 \documentclass[twocolumn]{aastex61}

\received{March 17, 2017}
\accepted{April 19, 2017}

\begin{document}
\title{Radii and mass-loss rates of type IIb supernova progenitors}

\correspondingauthor{Ryoma Ouchi}
\email{ouchi@kusastro.kyoto-u.ac.jp}
\affiliation{Department of Astronomy, Kyoto University, Kitashirakawa-Oiwake-cho, Sakyo-ku, Kyoto 606-8502, Japan}
%\author[0000-0002-0786-7307]{Greg J. Schwarz}
%\affil{American Astronomical Society \\
%2000 Florida Ave., NW, Suite 300 \\
%Washington, DC 20009-1231, USA}
\author{Ryoma Ouchi}
\affiliation{Department of Astronomy, Kyoto University, Kitashirakawa-Oiwake-cho, Sakyo-ku, Kyoto 606-8502, Japan}

\author{Keiichi Maeda}
\affiliation{Department of Astronomy, Kyoto University, Kitashirakawa-Oiwake-cho, Sakyo-ku, Kyoto 606-8502, Japan}

%\collaboration{(AAS Journals Data Scientists collaboration)}

%\author{Butler Burton}
%\affiliation{National Radio Astronomy Observatory}
%\affiliation{AAS Journals Associate Editor-in-Chief}
%\nocollaboration

%\altaffiliation{Creator of AASTeX v6.1}
%\collaboration{(LaTeX collaboration)}

%\affiliation{AAS Director of Publishing}
%\affiliation{American Astronomical Society \\
%2000 Florida Ave., NW, Suite 300 \\
%Washington, DC 20009-1231, USA}

%\author{Jeff Lewandowski}
%\affiliation{IOP Senior Publisher for the AAS Journals}
%\affiliation{IOP Publishing, Washington, DC 20005}



\begin{abstract}
    Several Type IIb supernovae (SNe IIb) have been extensively studied, both in terms of the progenitor radius and the mass-loss rate in the final centuries before the explosion. While the sample is still limited, evidence has been accumulating that the final mass-loss rate tends to be larger for a more extended progenitor, with the difference exceeding an order of magnitude between the more and less extended progenitors. The high mass-loss rates inferred for the more extended progenitors are not readily explained by a prescription commonly used for a single stellar wind. In this paper, we calculate a grid of binary evolution models. We show that the observational relation in the progenitor radii and mass-loss rates may be a consequence of non-conservative mass transfer in the final phase of progenitor evolution without fine tuning. Further, we find a possible link between SNe IIb and SNe IIn. The binary scenario for SNe IIb inevitably leads to a population of SN progenitors surrounded by dense circumstellar matter (CSM) due to extensive mass loss ($\dot{M} \gtrsim 10^{-4} M_{\odot} \mathrm{yr}^{-1}$) in the binary origin. About 4 \% of all observed SNe IIn are predicted to have dense CSM, produced by binary non-conservative mass transfer, whose observed characteristics are distinguishable from SNe IIn from other scenarios. Indeed, such SNe may be observationally dominated by systems experiencing huge mass loss in the final $10^{3}$ yr, leading to luminous SNe IIn or initially bright SNe IIP or IIL with a characteristics of SNe IIn in their early spectra. 

\end{abstract}




\section{Introduction}\label{sec:introduction}

Type IIb supernovae (SNe IIb) are characterized by hydrogen lines in their early phase spectra, which are gradually replaced by He lines at later phases \citep{1997ARA&A..35..309F}.
The progenitor of SNe IIb is believed to be a massive star which retains only a small amount of hydrogen ($\lesssim 1M_{\odot}$) in its outer layer at the time of the explosion. For the removal of the hydrogen layer, two scenarios have been considered. In one scenario, massive single stars ($\gtrsim 25M_{\odot}$) eject their outer layer via their strong stellar wind \citep{2012A&A...538L...8G, 2016MNRAS.455..112G}.  
The other scenario the binary interaction, i.e., a star in a binary system transfers most of its hydrogen-rich layer to its companion by Roche lobe overflow (RLOF) \citep{2009MNRAS.396.1699S}. The question as to which is the dominant evolutionary scenario for the production of SNe IIb progenitors is still open, but observations and theoretical models so far seem to favor the binary scenario \citep{2011MNRAS.412.1522S, 2012Sci...337..444S, 2014ApJ...793L..22F}.

Among SNe IIb, 1993J, 2008ax, 2011dh, and 2013df have been investigated in detail and have yielded abundant observational data, covering the long-term evolution from early to late phases at various wavelengths. Their progenitors (or strong candidates) have also been identified \citep{2004Natur.427..129M, 2011ApJ...739L..37M, 2014AJ....147...37V, 2015ApJ...811..147F}. In Table 1, characteristic properties of these progenitors are listed. Note the diversity they show in the HR diagram. They cover a large range in their radii, from $\sim 50R_{\odot}$ (blue supergiant : BSG) to $\sim 600R_{\odot}$ (yellow or red supergiant : YSG or RSG). Whether they represent two discrete groups or a continuous distribution is still uncertain.  
Various evolutionary models have been investigated for each progenitor that match both their location in the HR diagram and classification of their SN type. Binary evolution models seem to explain the observed features more naturally than single-star evolution models for most of them \citep{1994ApJ...429..300W, 2013ApJ...762...74B}. According to these works, the progenitor masses for these SNe are estimated to be in the range $12 \sim 18 M_{\odot}$.

In addition to information on the progenitors, the mass-loss rate just before the explosion contains important information about their evolutionary paths. The mass-loss property is reflected in the density of circumstellar matter (CSM), which has been studied by radio, X-ray, and optical observations in the late phase, through the signature of the SN-CSM interaction. \citet{2015ApJ...807...35M} have found a correlation between the progenitor radius and the average mass-loss rate shortly before the explosion thus derived, in which more extended progenitors ($\sim 600R_{\odot}$ ; e.g. 1993J, 2013df) have had a relatively large mass-loss rate before the explosion at a rate of $\dot{M}\sim 5\times 10^{-5}M_{\odot} \  \mathrm{yr^{-1}}$ \citep{1996ApJ...461..993F}, while less extended progenitors ($\sim 200R_{\odot}$ ; e.g. SN 2011dh) have had a moderate mass-loss rate ($\dot{M} \sim 3 \times 10^{-6}M_{\odot} \ \mathrm{yr^{-1}}$) \citep{2014ApJ...785...95M}. This tendency is supported by a larger sample of SNe without direct progenitor detection \citep{2016ApJ...818..111K}.

 For the less extended BSG or YSG progenitors like SN 2008ax or SN 2011dh, a mass-loss rate from a single stellar wind seems to be compatible with the observationally derived mass-loss rates. However, in the case of the more extended progenitors like SN 1993J or SN 2013df, which are also YSGs, a mass-loss rate under the commonly used prescription for a single stellar wind falls significantly short of the observationally derived values \citep{1988A&AS...72..259D}. It may still be possible that the extensive mass loss for the more extended progenitors can be explained solely by a single stellar wind, considering that the high mass-loss rates reaching as high as $\dot{M} \sim 10^{-4} M_{\odot} \mathrm{yr}^{-1}$ from some RSGs have been reported \citep{2005A&A...438..273V}. At the same time, this unusually extensive mass loss for the more extended progenitors might indicate an additional mass-loss mechanism related to binary evolution. Solving this problem may provide us with a key to understanding the evolutionary history of the progenitors of SNe IIb.

In the context of the binary scenario, the binary interaction, especially the non-conservative mass transfer, may be the origin of this additional mass loss  \citep{2011A&A...528A..16V}. Recently, \citet{2017arXiv170102089Y} showed that the mass-loss rates of the binary models for SNe IIb are consistent with the observationally derived values. They, however, did not discuss whether the relation between the progenitor radii and mass-loss rates are generally expected for different values of the initial mass ratios and mass accretion efficiency. Also, they did not discuss what kind of physical processes are involved in determining the mass-loss rate. In this paper, we investigate whether the apparent relation between progenitor size and mass-loss rate before the explosion can be explained by binary evolution models. We have found that this observed tendency can indeed be naturally explained by non-conservative mass transfer in the final phase. We also discuss the physical mechanisms that produce the relation.

We also suggest a possible link between binary evolution (and SNe IIb) and some Type IIn SNe (SNe IIn), which are characterized by narrow hydrogen emission lines in their spectra \citep{1997ARA&A..35..309F}. SNe IIn are believed to have dense CSM in the vicinity of the progenitors, which indicates extensive mass loss shortly before the explosion. The mass-loss rates have been estimated for many SNe IIn in various ways, and these values cover the range $10^{-4}$--$1 M_{\odot} \mathrm{yr}^{-1}$, often assuming a wind velocity of $\sim 100$ km/s \citep{2012ApJ...744...10K, 2013A&A...555A..10T, 2014MNRAS...439..2917M}. For a certain range of binary parameters, our binary models can have dense CSM comparable to these observationally derived values at the time of the explosion, produced by non-conservative mass transfer shortly before it.

This paper is structured as follows. In \S 2, we describe the method used for the calculation of the binary evolution models. In \S 3, we show the results of these models, focusing on the property of the progenitor and its relation to the mass-loss rate. In \S 4.1, we discuss how and why the binary model predicts that relation, by considering key physical processes in the binary evolution. In \S 4.2, possible evolutionary paths to SNe IIn through binary evolution are discussed. Our results are summarized in \S 5.

\section{Method} \label{sec:method}

We use MESA\footnote[1]{'http://mesa.sourceforge.net/'} for the calculations of the binary evolution. MESA is a one-dimensional stellar evolution code that uses adaptive mesh refinement and adaptive time stepping.
See \citet{2011ApJS..192....3P, 2013ApJS..208....4P, 2015ApJS..220...15P} for details. In this section, we briefly describe the key parameters in our calculations.

 We assume a solar metallicity of $Z=0.02$ for all the models. Convection is modeled using the mixing length theory of \citet{1965ApJ...142..841H}, adopting the Ledoux criterion. The mixing length parameter is set to be $\alpha=2.0$. Semiconvection is modeled following \citet{1985A&A...145..179L} with an efficiency parameter $\alpha_{sc}=1.0$ \citep{2010ApJ...725..940Y}.
Thermohaline mixing is not included in our simulations. As for convective overshooting, we follow the diffusive approach of \citet{2000A&A...360..952H}, with $f=0.018$ and $f_0=0.002$, meaning that the overshooting extends to $\sim 0.016H_P$ from the convective boundaries ($H_P$ is the scale height near the convective boundaries). We adopt the overshooting for the convective core during hydrogen-burning and the convective hydrogen-burning shell, and also for the convective core and shell where no significant burning takes place. We use the `Dutch' scheme for the stellar wind, with a scaling factor of 1.0. The `Dutch' wind scheme in MESA combines results from several papers. Specifically, when $T_{\mathrm{eff}} > 10^{4}$K and the surface mass fraction of hydrogen is greater than 0.4, we then use that of \citet{2001A&A...369..574V}, and when $T_{\mathrm{eff}} > 10^{4}$K and the surface mass fraction of hydrogen is less than 0.4, we use that of \citet{2000A&A...360..227N}. Then, in the case when $T_{\mathrm{eff}} < 10^{4}$K, we use the wind scheme of \citet{1988A&AS...72..259D}. When the primary star fills its Roche lobe, we implicitly compute the mass transfer rate following the scheme of \citet{1990A&A...236..385K}.

 Some previous binary evolution calculations, including stellar rotation have shown that the mass accretion from the primary can bring the secondary close to the critical rotation, which is then likely to enhance the wind. This enables the mass transfer to be highly non-conservative, especially for binaries with relatively large orbital period ($\gtrsim$ 10 days). This is usually expected for Type IIb binary progenitor models \citep{2003astro.ph..2232L, 2005A&A...435.1013P}. So, in this paper, we consider only non-conservative mass transfer and assume the efficiency of mass accretion $f$ to be 0.5 or 0.0, which is kept constant throughout the calculations. Here, the mass accretion efficiency $f$ denotes the fraction of the mass transferred by the primary which accretes onto the secondary. We assume that the matter ejected by non-conservative mass transfer has a specific angular momentum equal to that of the accreting star.

In some of our models, it happens that both stars fill their own Roche lobes at the same time. In this case, it is likely that the system enters the common envelope phase, which MESA currently cannot deal with. In this case we stop the calculation. Furthermore, in several models, after the major mass transfer by the primary has occurred, the secondary completes the main-sequence stage and expands to become a giant. This results in the RLOF of the secondary, before the primary star explodes. Also in this case we stop the calculation.

In order to compare the model outcomes with the observationally derived mass-loss rates before the explosion, we calculate the final mass-loss rate for each model as follows. We pick up a model snapshot at about 1000 yr before the end of the calculation, and compute the difference between the total mass of the two stars at this epoch and that at the end of the calculation. This is then divided by the time interval between the two phases. Throughout this paper, we denote this quantity as $\dot{M}_{1000}$.

\begin{table*}[tbp]
  \begin{center}
\begin{tabular}{|l||c|c|c|c|r|} \hline
  &  log($T_{\mathrm{eff}}$(K)) &  log($L/L_{\odot}$)   &   Radius($R_{\odot}$)  & $\dot{M}(M_{\odot} \ \mathrm{yr^{-1}})$    \\ \hline \hline
  1993J & 3.63 $\pm$ 0.05 & 5.1 $\pm$ 0.3  & $\sim$ 600      & (2--6)$\times 10^{-5}$    \\
  2008ax & 3.9--4.3 & 4.4--5.25   & $\sim$ 50 & $6.5\times 10^{-6} $ \\
  2011dh & 3.76--3.80 & 4.92$\pm$ 0.20 & $\sim$ 200    & $3\times 10^{-6}$ \\
  2013df &  3.62--3.64  & 4.94$\pm$ 0.06 & $\sim$ 600    & (5.4$\pm$3.2)$\times 10^{-5}$    \\ \hline
\end{tabular}
\caption{Properties of the progenitors of Well-studied Type IIb SNe. For information on the HR diagram and the progenitor radius, the date are taken from \citet{2004Natur.427..129M} for SN 1993J, \citet{2015ApJ...811..147F} for SN 2008ax, \citet{2011ApJ...739L..37M} for SN 2011dh, and \citet{2014AJ....147...37V} for SN 2013df. The data of the mass-loss rate are taken from \citet{1996ApJ...461..993F} for SN 1993J, \citet{2010ApJ...711L..40C} for SN 2008ax, \citet{2014ApJ...785...95M} for SN 2011dh, and from \citet{2015ApJ...807...35M} for SN 2013df.}
  \end{center}
\end{table*}  

The initial primary star mass is fixed to be $16M_{\odot}$, so as to be roughly consistent with the luminosity of the detected progenitors. For the secondary star mass, we consider different values of the initial mass ratio \footnote[2]{The subscripts 1 and 2 express the primary and the secondary respectively. The primary in this paper refers to the more massive star at the beginning of the calculation.} $q=M_2/M_1$, simulating the models with $q=0.6, 0.8, 0.95$. We note that, as shown by \citet{2012Sci...337..444S}, the initial mass ratios implied for Galactic O stars appear to be uniformly distributed. Therefore, systems with $q \lesssim 0.6$ do exist. However, because of the small mass ratio, these binaries are expected to have a common envelope phase (Table 2, Table 3), once they begin RLOF. Further evolution of those models requires complicated considerations, which are beyond the scope of this paper. We also note that the properties of the Type IIb progenitors in Table 1 are explained by the relatively high mass ratio ($q \gtrsim 0.6$) \citep{2009MNRAS.396.1699S, 2013ApJ...762...74B, 2015ApJ...811..147F}, which may indicate that these Type IIb progenitors mainly evolve from binaries with high initial mass ratio ($q \gtrsim 0.6$).

We perform binary evolution simulations for different values of the initial period $P=$5, 25, 50, 200, 600, 800, 1200, 1600, 1800, 1950 and 2200 days. We follow the evolution of both stars, from the zero-age main sequence until the mass fraction of carbon at the primary's center falls below $1.0\times 10^{-6}$. This corresponds to the carbon-shell burning phase and it takes only a few years until the explosion from that point.

We assume that models with final envelope mass more than $1M_{\odot}$ explode as SNe IIP or SNe IIL, while those with envelope mass less than $0.01M_{\odot}$ explode as SNe Ib or Ic. Models whose envelope mass is between $0.01M_{\odot}$ and $1M_{\odot}$ are adopted as SN IIb progenitors.

\section{Results} \label{sec:results}

Tables 2 and 3 show the final properties of each model, including the average mass-loss rates before the explosion ($\dot{M}_{1000}$), for an accretion efficiency of $f=0.5$ and 0.0, respectively. Fig. 1 shows the final hydrogen-rich envelope mass of the progenitor models as a function of the initial orbital period. It is clear that the differences in the initial mass ratio and the accretion efficiency do not sensitively affect the final content of the envelope mass, while the initial period does. Furthermore, there is a tendency that the primary star retains a larger amount of the hydrogen-rich envelope at the time of the explosion for the larger initial period. The models with $P=5$ days lose all of their hydrogen-rich envelope and become SNe Ib/c. Following our criterion, the initial period of 1000 days separates SN IIP/IIL and SN IIb progenitors. Thus the models with initial period in the range 10 days $\lesssim P \lesssim$ 1000 days may become SNe IIb.

\begin{figure}[htbp]
 \begin{minipage}{0.5\hsize}
%%\hspace*{10mm}
   \begin{center}
      \includegraphics[width=80mm]{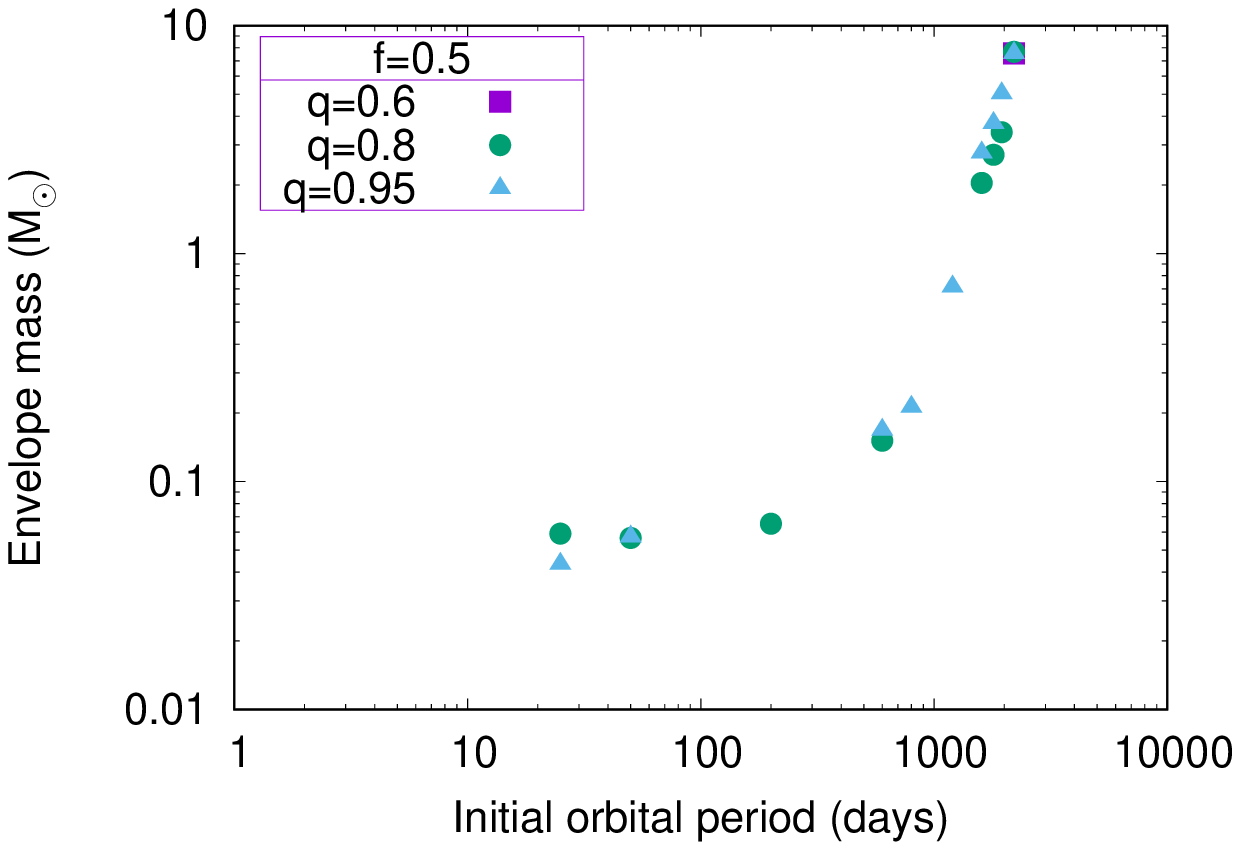}
   \end{center}
  \end{minipage}
  %%\hspace*{10mm}
  \begin{minipage}{0.5\hsize}
    \begin{center}
      \includegraphics[width=80mm]{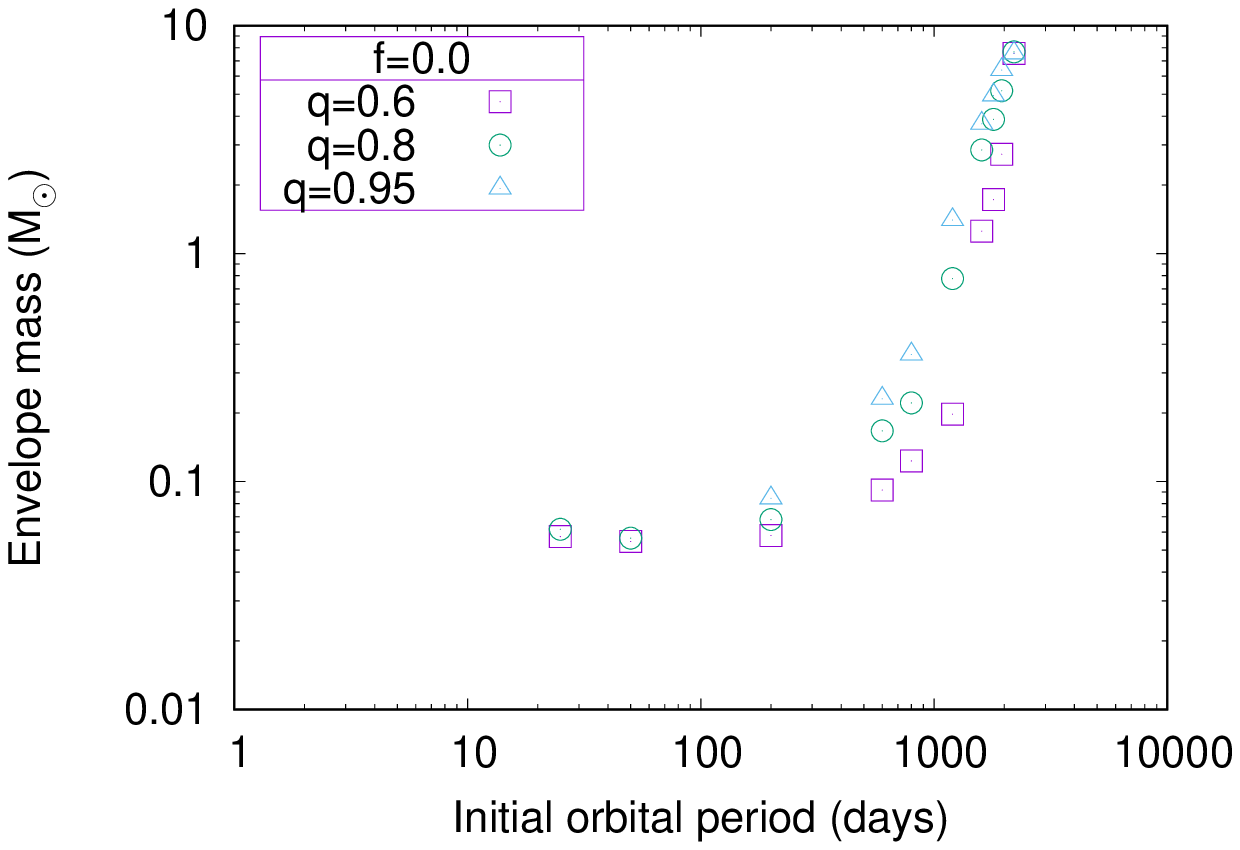}
   \end{center}
  \end{minipage}
 %% \vspace*{1.4cm} 
  \caption{Final envelope mass of the progenitor as a function of the initial orbital period $P$. The left and right panels show the models with the accretion efficiency of $f=0.5$ and $f=0.0$, respectively. The models with different initial mass ratios are shown by different symbols/colors.}
\end{figure}

Fig. 2 shows the final locations of the SN IIb progenitor models on the HR diagram (the SN II and SN Ib/c progenitor models are not included in the figure). Note that our models cover the range of properties that the detected progenitors show in the HR diagram. From this figure, together with Table 2 and Table 3, it is clear that the models with a smaller initial period are located at the left side of the HR diagram, i.e., they are more compact. Furthermore, the final locations on the HR diagram are not sensitively dependent on the initial mass ratio and the accretion efficiency compared with the initial period.

\begin{figure}[htbp]
%%  \hspace*{10mm}
 \begin{center}
   \includegraphics[width=80mm]{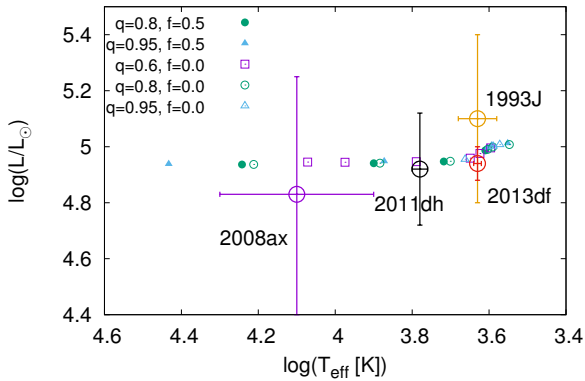}
  %% \vspace*{1.0cm}
   \end{center}
  \caption{Final locations of the SN IIb progenitor models, together with the the observed values for the detected progenitors, on the HR diagram.}
\end{figure}

Next, we plot progenitor radius (hereafter $R$) versus $\dot{M}_{1000}$ in Fig. 3, together with the observed values for the well-studied progenitors. We find that the binary models do predict the relation between the progenitor radius and the mass loss, without any fine-tuning. Namely, the more extended progenitors have about an order of magnitude higher mass-loss rates before the explosion than the less extended ones. The outliers in the relation, i.e., those with $R\sim 890R_{\odot}$ and $\dot{M} \sim$ (5--7) $\times$ $10^{-6}M_{\odot} \ \mathrm{yr}^{-1}$ are the models with $P = 2200$ days. The Roche lobe radii (hereafter $R_{\mathrm{rl}}$) of these models are too large for the primary stars to start the mass transfer, and the final mass-loss rate is mostly determined by the stellar wind. They may explode as SNe IIP.

Our models leading to the less extended progenitors (e.g., SN 2008ax, SN 2011dh) reproduce the absolute values of the mass-loss rates observed for these progenitors fairly well. However, for the more extended progenitors like 1993J and 2013df, the values of the mass-loss rates found in our models are slightly lower than the observations indicate. Nevertheless, given the intrinsic uncertainties in the stellar evolution calculations and the measurement of the mass-loss rates, further investigation of this issue is beyond the scope of this paper.

\begin{figure}[htbp]
  %%\hspace*{10mm}
  \centering
  \includegraphics[width=80mm]{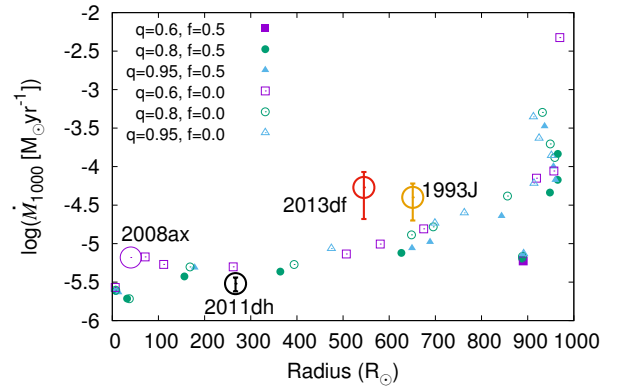}
  %%\vspace*{0.8cm}
  \caption{Radius versus mass-loss rate for all the models, together with the observationally derived values.}
\end{figure}

\section{Discussion} \label{sec:discussion}

\subsection{The radius versus mass-loss relation}

In all models with an initial period of $P = 5$ days, and most of the models with an initial period of $P = 25$ days, there is no RLOF after He-burning, so the final mass-loss rate is determined by the stellar wind, which has a value of $\dot{M} \sim 10^{-6} M_{\odot} \  \mathrm{yr}^{-1}$. The final mass-loss rates of the models with $P = 2200$ days are also determined by the stellar wind (Section 3). In all other models, the primary star experiences RLOF after exhaustion of the He fuel, and the RLOF continues until the explosion.
For the less extended progenitors whose radius is $R \lesssim 200R_{\odot}$, the mass loss via non-conservative mass transfer is a few $\times 10^{-6} M_{\odot}  \ \mathrm{yr}^{-1}$, which is comparable to or a little higher than the stellar wind (see Table 2, 3). For the more extended progenitrs, the mass transfer rates reach $\sim 10^{-5} M_{\odot}  \ \mathrm{yr}^{-1}$, or even larger, which far exceed the mass-loss rate predicted by the stellar wind. In this case, the mass-loss rate is mostly determined by the RLOF mass transfer rate ($\dot{M}_{\mathrm{tr}}$) in the final phase. In this subsection, we discuss why a more extended progenitor has a higher mass transfer rate in the final phase in our binary models.

Omitting the models in which there is no, or a negligible amount of RLOF in the final phase, such as those with $P=$5 or 2200 days, the mass transfer history is divided into two classes, i.e. Case BB, and Case C \citep{2015PASA...32...15Y}. For models with $P \lesssim 1400$ days, the Roche lobe radius is relatively small, so that the first mass transfer occurs when the primary is in the hydrogen-shell-burning phase. At this time, with the hydrogen-rich envelope almost intact, this mass transfer is usually thermally or dynamically unstable and occurs on a very short timescale. Then, as He-burning sets in, the primary shrinks, and the binary becomes detached. After exhaustion of the He fuel, the primary expands again while the carbon-oxygen core contracts and thus the second mass transfer begins. This mass transfer is moderate, being stable both dynamically and thermally, and continues until the time of the explosion.

On the other hand, for models with $P \gtrsim 1400$ days, the Roche lobe radius is relatively large and mass transfer begins only after exhaustion of the He fuel and continues until the explosion (case C). In this case, at the beginning of RLOF, the mass transfer is unstable and intense for a short period, reaching $\gtrsim (10^{-3}$--$10^{-2})M_{\odot} \ \mathrm{yr}^{-1}$ initially. Shortly after this phase, a moderate and stable mass transfer takes place as driven by the expansion due to the core-evolution. For a certain range of the initial period, this short intensive mass transfer phase occurs very shortly before the explosion, likely leading to a shell-like, dense CSM located near the progenitor at the time of the explosion. We will discuss this issue further in \S 4.2.2. 

In Fig. 4, the progenitor radius and the Roche lobe radius (left), and the mass transfer rate during the last $1.5\times 10^4$ years (right) for a typical model in the case BB mass transfer (No. 8 in Table 2) are shown. The rise in radius until about $1.2105\times 10^7 \ \mathrm{yr}$ corresponds to the carbon-oxygen core contraction (He-shell burning) phase, and the carbon burning starts at the dip in the evolution of radius at $\sim 1.2105\times 10^7 \ \mathrm{yr}$. Carbon burning continues for $\sim 5\times 10^{3}$ yr (until the next dip), and this is then followed by carbon-shell burning, with a rapid increase in radius.

\begin{figure}[htbp]
%%\centering
  \begin{minipage}{0.5\hsize}
%%\hspace*{10mm}
   \begin{center}
      \includegraphics[width=80mm]{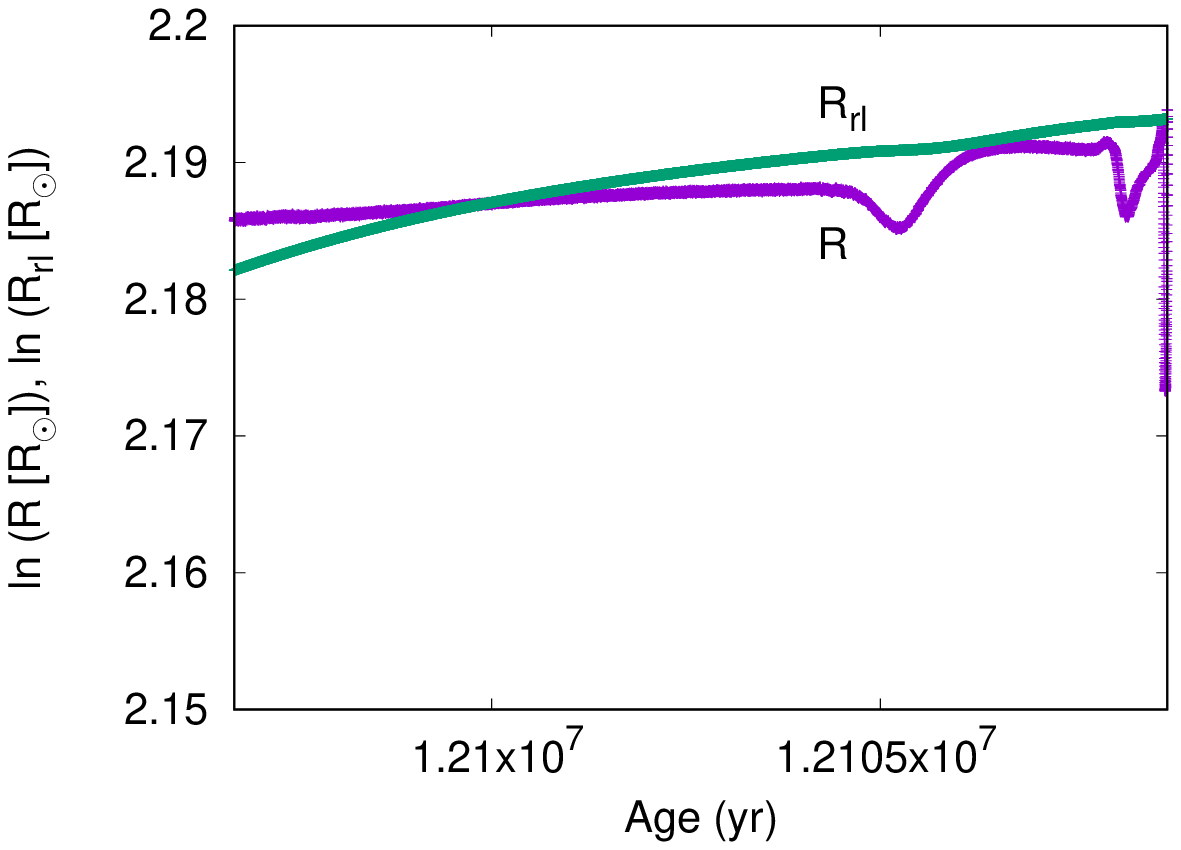}
   \end{center}
  \end{minipage}
  \begin{minipage}{0.5\hsize}
%%\hspace*{10mm}
    \begin{center}
      \includegraphics[width=80mm]{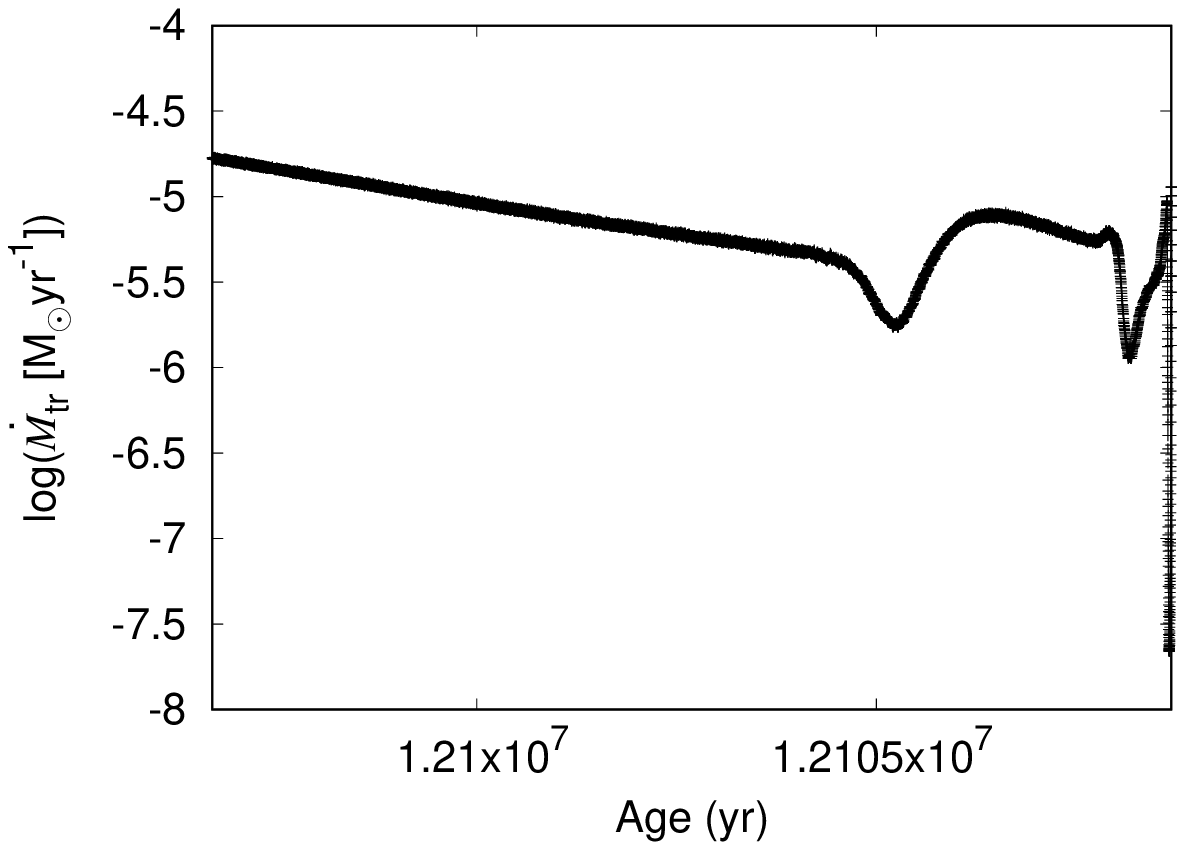}
    \end{center}
  \end{minipage}
  %%\vspace*{1.2cm}
  \caption{Left: time evolution of the stellar radius ($R$) and Roche lobe radius ($R_{\mathrm{rl}}$) during the last $1.5\times 10^4$ years of the primary star's evolution. The purple thick line and green thin line represent the stellar radius and the Roche lobe radius, respectively. Right: time evolution of the mass transfer rate ($M_{\mathrm{tr}}$) during the same epoch as the left panel. Both are plotted for the model No. 8 in Table 2.}
\end{figure}

Note that, before the ignition of carbon burning, the mass transfer is driven by the expansion due to the carbon-oxygen core contraction, and the radius remains almost equal to the Roche lobe radius. Then, once carbon burning sets in, the radius changes rapidly on a short timescale. The rapid change in radius during this phase causes the mass transfer rate to fluctuate by some factors around the value at the beginning of carbon burning. The final evolution in the last $\sim 10^3$ yr toward the explosion corresponds to this fluctuation phase, and it is difficult to estimate the mass transfer rate during this phase by simple analysis.

In the case of the stable mass transfer, the mass transfer rate does not change significantly after carbon ignition (Fig. 4). In Fig. 5, we compare $\dot{M}_{1000}$ of the models with $f$=0.0, which is approximately equal to the average mass transfer rate in the last $1 \times 10^3$ yr, with the mass transfer rate at $6 \times 10^3$ yr before the explosion, which we denote as $\dot{M}_{tr, 6000}$. This epoch, $6 \times 10^3$ yr before the explosion, corresponds to the carbon-ignition phase. These two values are closely connected, distributed tightly on the line $\dot{M}_{tr, 6000}=\dot{M}_{1000}$. This supports the idea that the mass transfer rate does not change significantly after carbon ignition. Although some models with relatively low mass transfer rate slightly deviate from the line $\dot{M}_{tr, 6000}=\dot{M}_{1000}$, this is because the contribution of the stellar wind to the mass loss is not negligible at such a low rate. In addition, there are some models with high mass-loss rates ($\gtrsim 10^{-4} M_{\odot} \ \mathrm{yr}^{-1}$) which deviate from the line significantly. This is because an unstable RLOF phase is involved during the last $6 \times 10^3 \mathrm{yrs}$. We discuss this issue in \S 4.2.2.

Making use of the fact that the mass transfer rate does not change significantly after carbon ignition for stable mass transfer, we hereby approximate the average mass transfer rate in the last $10^3$ yr, to compare with the observations, by the mass transfer rate at the ignition of carbon burning as analyzed by simple arguments. In this way, we analyze what mechanisms determine the mass transfer rate at the ignition of carbon burning.

\begin{figure}[tbp]
  \hspace*{3mm}
  \begin{center}   
    \includegraphics[width=100mm]{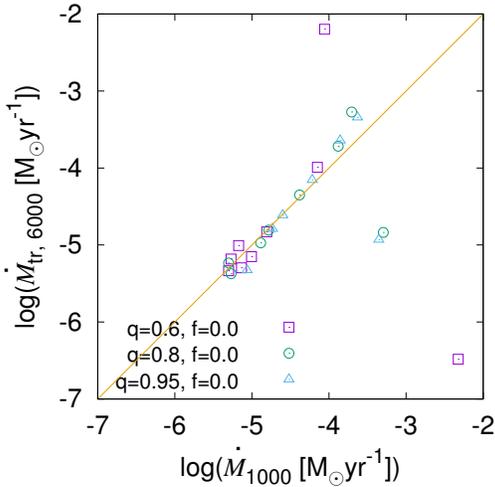}
    %%\vspace*{1.0cm}
\end{center}
    \caption{$\dot{M}_{1000}$ is compared to the mass transfer rate  at $6\times 10^3$ yrs before the explosion ($\dot{M}_{tr, 6000}$), which corresponds to the ignition of carbon burning. Shown are the models with $f=0.0$.}
\end{figure}

During the stable mass transfer driven by carbon-oxygen core contraction, the mass transfer rate can be approximated as follows \citep{1997A&A...327..620S, 2015ebss.book..179I};

\begin{eqnarray}
-\dot{M}_{\mathrm{tr}}&=&\frac{M_{\mathrm{env}}}{\zeta_{\mathrm{eq, env}}-\zeta_{\mathrm{L}}\frac{M_{\mathrm{env}}}{M}}\left.\Bigl(\frac{\partial \mathrm{ln}R}{\partial t}-\frac{\partial \mathrm{ln}R_{\mathrm{rl}}}{\partial t} \Bigr)\right|_{\dot{M}=0} \ .
\end{eqnarray}

Here, $M_{\mathrm{env}}$ and $M$ represent the hydrogen-rich envelope mass and the total mass of the donor star, respectively. This relation can be derived from

\begin{eqnarray}
\frac{d\mathrm{ln}R}{dt}=\zeta_{\mathrm{eq, env}}\frac{d\mathrm{ln}M_{\mathrm{env}}}{dt}+\left.\frac{\partial \mathrm{ln}R}{\partial t}\right|_{\dot{M}=0}  \   , \mathrm{and}
  \end{eqnarray}

\begin{eqnarray}
\frac{d\mathrm{ln}R_{\mathrm{rl}}}{dt}=\zeta_{\mathrm{L}}\frac{d\mathrm{ln}M}{dt}+\left.\frac{\partial \mathrm{ln}R_{\mathrm{rl}}}{\partial t}\right|_{\dot{M}=0}    \ ,
  \end{eqnarray}

by assuming that the equation $R=R_{\mathrm{rl}}$ is fulfilled all the time, which is approximately correct during the phase we are focusing on here (Fig. 4). Here, $\zeta_{\mathrm{eq, env}}$ expresses the change in donor radius in response to the reduction of the envelope mass, assuming that this mass loss occurs slowly enough to keep the donor in thermal equilibrium. $\zeta_{\mathrm{L}}$ is  the change in the Roche lobe radius of the donor in response to the mass transfer. They are defined as follows:

\begin{eqnarray}
  \zeta_{\mathrm{eq, env}}&=&\Bigl(\frac{\partial \mathrm{ln}R}{\partial \mathrm{ln}M_{\mathrm{env}}}\Bigr)_{\mathrm{eq}} \    , \mathrm{and} \\
  \zeta_{\mathrm{L}}&=&\frac{\partial \mathrm{ln}R_{\mathrm{rl}}}{\partial \mathrm{ln}M}   \  .
\end{eqnarray}

The second term in brackets of equation (1) expresses the angular momentum loss due to gravitational wave emission, and this is negligible compared to the first term in our binary models. Then, we can rewrite equation (1) as follows:
\begin{eqnarray}
-\dot{M}_{\mathrm{tr}}&=&\frac{M_{\mathrm{env}}}{\zeta_{\mathrm{eq, env}}-\zeta_{\mathrm{L}}\frac{M_{\mathrm{env}}}{M}}\left.\Bigl(\frac{\partial \mathrm{ln}R}{\partial t}\Bigr)\right|_{\dot{M}=0}    . 
\end{eqnarray}
Physically, the term of $M_{\mathrm{env}} \times \left.\Bigl(\frac{\partial \mathrm{ln}R}{\partial t}\Bigr)\right|_{\dot{M}=0}$ means that as the donor expands considerably ($\Delta R \sim R$) in a timescale of $\left.\Bigl(\frac{\partial \mathrm{ln}R}{\partial t}\Bigr)\right|^{-1}_{\dot{M}=0}$, the whole envelope mass is lost. The term of $\zeta_{\mathrm{eq, env}}$ represents the effect of change in the donor radius in the equilibrium state due to the mass loss, while $\zeta_{\mathrm{L}}$ represents the change in the Roche lobe radius due to the mass transfer. In sum, five factors ($\zeta_{\mathrm{eq, env}}, \zeta_{\mathrm{L}}, M_{\mathrm{env}}, M, \left.\frac{\partial \mathrm{ln}R}{\partial t}\right|_{\dot{M}=0}$) determine the mass transfer rate. We consider these terms one by one below.

First, we consider $\zeta_{\mathrm{eq, env}}$. In order to estimate this value, we calculate a series of single-star models having the same physical condition as the primary star in our binary models, except that here we remove the envelope mass artificially at a high rate ($\dot{M}\sim 10^{-3}M_{\odot}  \ \mathrm{yr}^{-1}$) during He-burning and turn off the stellar wind of the primary star until exhaustion of the carbon fuel. This procedure allows us to construct a series of the primary models with different amounts of the hydrogen-rich envelope under thermal equilibrium. By extracting the radius and the envelope mass at the beginning of carbon burning of these single-star models, we create a relation between the envelope mass and the radius in its equilibrium state. We then derive $\zeta_{\mathrm{eq, env}}$ by differentiating the curve. Fig. 6 shows such a plot. Although this figure is constructed by evolving the single stars, the plot is applicable to our primary star at the same evolutionary stage, because the physical conditions assumed are the same.

From Fig. 6, we see that the radius in the equilibrium state increases rather quickly with increasing envelope mass as long as ln ($M_{\mathrm{env}} [M_{\odot}]$) $\lesssim$ -2, while beyond this point, the radius is almost constant as a function of the envelope mass. This sudden change of the shape of the curve at this point is due to the low surface temperature and thus the development of the convective layer in the envelope, when ln ($M_{\mathrm{env}} [M_{\odot}]$) $\gtrsim$ -2. As shown in Fig. 6, the radius in complete equilibrium increases with the increase of the envelope mass when the envelope is radiative (ln ($M_{\mathrm{env}} [M_{\odot}]$) $\lesssim$ -2). This behavior can be explained approximately by an analytical argument, by applying the approach of \citet{1961ApJ...133..764C} to the situation under consideration (see Appendix A for details).   
For an envelope mass beyond ln ($M_{\mathrm{env}} [M_{\odot}]$) $\sim$ -2 in the convective regime, further increase of the radius is suppressed by the existence of the Hayashi line, which keeps the radius almost constant.
 Therefore, $\zeta_{\mathrm{eq, env}}$ is close to zero for more extended progenitors ($M_{\mathrm{env}}\gtrsim 0.8M_{\odot}$), while $\zeta_{\mathrm{eq, env}}\gtrsim 1$ for less extended progenitors ($M_{\mathrm{env}} \lesssim$ 0.8$M_{\odot}$). 
\begin{figure}[htbp]
%%\hspace*{10mm}
  \centering
  \includegraphics[width=80mm]{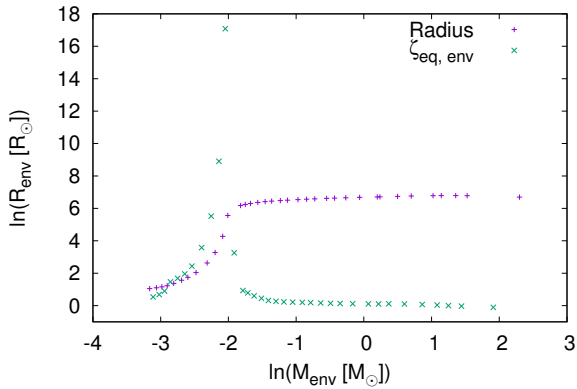}
 %% \vspace*{0.85cm}
  \caption{Equilibrium radius (purple) and $\zeta_{\mathrm{eq, env}}$ (green) as a function of the hydrogen-rich envelope mass at the ignition of carbon burning.}
\end{figure}

Next, we consider $\zeta_{\mathrm{L}}$. After some calculations, using the approximation formula by \citet{1983ApJ...268..368E}, $\zeta_{\mathrm{L}}$ can be written as follows \citep{1997A&A...327..620S}:

\begin{eqnarray}
\zeta_{\mathrm{L}}&=&\frac{\partial \mathrm{ln}R_{\mathrm{rl}}}{\partial \mathrm{ln}M}  \nonumber \\
&=&\frac{\partial \mathrm{ln}a}{\partial \mathrm{ln}M}+\frac{\partial \mathrm{ln}(R_{\mathrm{rl}}/a)}{\partial \mathrm{ln}q}\frac{\partial \mathrm{ln}q}{\partial \mathrm{ln}M}  \nonumber \\
  &=&\biggl(1+\frac{1-\beta}{q}\biggr)\biggl(-\frac{4}{3}-\frac{1}{1+q}+\frac{5-3\beta}{1-\beta+q} \nonumber \\ &&-\frac{1}{3q^{1/3}}\frac{1.2q^{1/3}+\frac{q}{1+q^{1/3}}}{0.6+q^{2/3}\mathrm{ln}(1+q^{-1/3})}
  \biggr) \ .
\end{eqnarray}

Here, we define $\beta\equiv 1-f$, as the fraction of the transferred material that is lost from the system. We plot $\zeta_{\mathrm{L}}$ in Fig. 7 for three values of the mass accretion efficiency. For a large mass ratio ($q \gg 1$), which is often expected in the final stage of the binary evolution we are considering, $\zeta_{\mathrm{L}}$ converges to $\sim -1.5$. This behavior is explained in an analytical way, and is described in Appendix B. 
\begin{figure}[tbp]
  %%\hspace*{10mm}
  \begin{center}   
    \includegraphics[width=80mm]{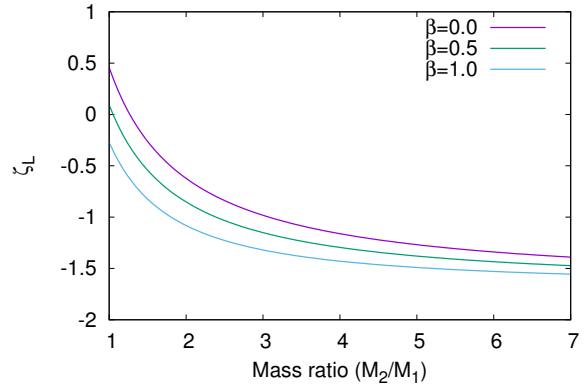}
    \vspace*{0.85cm}
    \caption{Value of $\zeta_{\mathrm{L}}$ as a function of the mass ratio, for different value of $\beta = M_2/M_1$.}
  \end{center}
\end{figure}

Finally, in Fig. 8, we plot $\left.\frac{\partial \mathrm{ln}R}{\partial t}\right|_{\dot{M}=0}$ as a function of the envelope mass at the ignition of carbon burning. In calculating this value, we use the same single-star models as those used to create Fig. 6. We pick up two phases: $6 \times 10^3$ yr before the end of the calculation (which is around the ignition of carbon burning) and $1 \times 10^3$ yr before that point. We then calculate the difference in ln $R$ between these two phases, which is then divided by the time interval between them ($\sim 1 \times 10^3$ yr). Fig. 8 shows that stars with a small amount of the hydrogen-rich envelope (ln $M_{\mathrm{env}}(M_{\odot}) \lesssim -2 $) have a large value of $\left.\frac{\partial \mathrm{ln}R}{\partial t}\right|_{\dot{M}=0} (\sim 1 \times 10^{-4} \ \mathrm{yr}^{-1}$), while those with a large amount of the envelope (ln $M_{\mathrm{env}}(M_{\odot}) \gtrsim -1 $) have a small value of $\left.\frac{\partial \mathrm{ln}R}{\partial t}\right|_{\dot{M}=0} (\sim 1 \times 10^{-5} \mathrm{yr}^{-1}$). The rapid decrease of the expansion rate $\left.\frac{\partial \mathrm{ln}R}{\partial t}\right|_{\dot{M}=0}$ beyond ln $M_{\mathrm{env}}(M_{\odot}) \sim -2$ is because the convective layer develops from that point and the expansion of the radius is suppressed by the Hayashi line.

\begin{figure}[htbp]
  %%\hspace*{10mm}
  \centering
  \includegraphics[width=80mm]{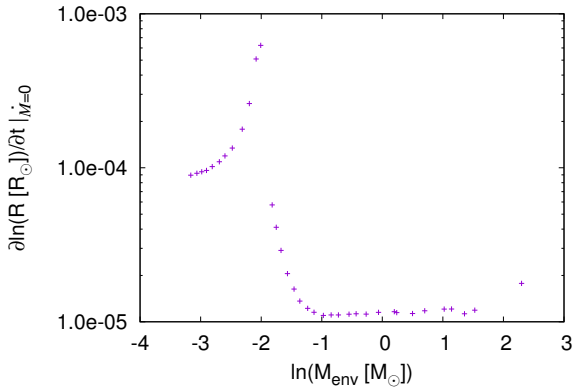}
  %%\vspace*{0.85cm}
  \caption{Expansion rate of the donor, assuming there is no mass loss, as a function of the envelope mass at the ignition of carbon burning.}
\end{figure}

From Fig. 6--8 and the discussions so far, a simple picture can be derived leading to the relation between the radius and mass-loss rate (Fig. 3). First, let us consider the less extended progenitors ($R \lesssim 300R_{\odot}$). These have a hydrogen-rich envelope as small as
$M_{\mathrm{env}}\sim 0.06 M_{\odot}$ (Tables 2 and 3), which then gives $\zeta_{\mathrm{eq, env}} \sim 1.5$ (Fig. 6). For the less extended progenitors, $\zeta_{\mathrm{L}} \sim -1$ (Fig.7, Tables 2 and 3), while $\frac{M_{\mathrm{env}}}{M} \sim 10^{-2}$. Therefore, $\zeta_{\mathrm{eq, env}}-\zeta_{\mathrm{L}}\frac{M_{\mathrm{env}}}{M} \sim \zeta_{\mathrm{eq, env}} \sim 1.5$. Recalling that for the less extended stars, $\left.\frac{\partial \mathrm{ln}R}{\partial t}\right|_{\dot{M}=0} \sim 1 \times 10^{-4} \ \mathrm{yr}^{-1}$ (Fig.8), the mass transfer rate at carbon ignition is estimated as follows:

\begin{eqnarray}
-\dot{M}_{RLOF}&=&\frac{M_{\mathrm{env}}}{\zeta_{\mathrm{eq, env}}-\zeta_{\mathrm{L}}\frac{M_{\mathrm{env}}}{M}}\left.\Bigl(\frac{\partial \mathrm{ln}R}{\partial t}\Bigr)\right|_{\dot{M}=0}  \nonumber \\
&\sim& \frac{0.06 M_{\odot}}{1.5}\times 1\times 10^{-4} \mathrm{yr}^{-1} \nonumber \\
&\sim& 4\times 10^{-6} M_{\odot} \ \mathrm{yr}^{-1} \ .
\end{eqnarray}
This value, taking the contribution from the stellar wind into account, is consistent with that extracted from the binary evolution calculations (Table 3).

Next, we consider a representative case among the more extended progenitors, with $R \sim 900 R_{\odot}$ having a hydrogen-rich envelope mass of $M_{\mathrm{env}}\sim 1 M_{\odot}$ (Table2 and 3). From the value of $M_{\mathrm{env}}$ and Fig. 6, we derive $\zeta_{\mathrm{eq, env}} \sim 0.1$. For these progenitors, $\zeta_{\mathrm{L}} \sim -1$ (Fig.7), while $\frac{M_{\mathrm{env}}}{M} \sim 0.15$. Therefore, $\zeta_{\mathrm{eq, env}}-\zeta_{\mathrm{L}}\frac{M_{\mathrm{env}}}{M} \sim 0.25$. Recalling that for the more extended stars, $\left.\frac{\partial \mathrm{ln}R}{\partial t}\right|_{\dot{M}=0} \sim 1 \times 10^{-5} \mathrm{yr}^{-1}$ (Fig.8), the mass transfer rate at carbon ignition is estimated as follows:   
\begin{eqnarray}
-\dot{M}_{RLOF}&=&\frac{M_{\mathrm{env}}}{\zeta_{\mathrm{eq, env}}-\zeta_{\mathrm{L}}\frac{M_{\mathrm{env}}}{M}}\left.\Bigl(\frac{\partial \mathrm{ln}R}{\partial t}\Bigr)\right|_{\dot{M}=0}  \nonumber \\
&\sim& \frac{1 M_{\odot}}{0.25}\times 1\times 10^{-5} \mathrm{yr}^{-1} \nonumber \\
&\sim& 4 \times 10^{-5} M_{\odot} \ \mathrm{yr}^{-1} \ .
\end{eqnarray}
This value is consistent with that extracted from the binary evolution calculations (Table 3).

 In summary, the difference between the mass-loss rate of less extended progenitors and that of more extended progenitors can be explained as follows.
Firstly, the value of $\left.\frac{\partial \mathrm{ln}R}{\partial t}\right|_{\dot{M}=0}$ is an order of magnitude lower for more extended progenitors than less extended ones. This alone is contrary to the observed tendency. However, the envelope mass $M_{\mathrm{env}}$ is about an order of magnitude larger for more extended progenitors than less extended ones. Furthermore, the value of $\frac{1}{\zeta_{\mathrm{eq, env}}-\zeta_{\mathrm{L}} \frac{M_{\mathrm{env}}}{M}}$ is also about an order of magnitude larger for more extended progenitors than less extended ones, mainly because $\zeta_{\mathrm{eq, env}}$ is an order of magnitude smaller for more extended ones. Therefore, these factors ($M_{\mathrm{env}}$ and $\zeta$-terms) cause the mass transfer rate to be two orders of magnitude higher for more extended progenitors than less extended ones. Combined with the effect of $\left.\frac{\partial \mathrm{ln}R}{\partial t}\right|_{\dot{M}=0}$, the mass transfer rate of more extended progenitors should be an order of magnitude higher.

Thus, the simple intuitive estimation of mass transfer rate by the $M_{\mathrm{env}} \times \left.\frac{\partial \mathrm{ln}R}{\partial t}\right|_{\dot{M}=0}$ is not enough to recover the relation in Fig. 3, and the $\zeta$ terms (especially $\zeta_{\mathrm{eq, env}}$) also play a key role in producing the relation.

\subsection{implications for Type IIn SNe}

\subsubsection{A rate of Type IIn SNe through the binary evolution}
As indicated in Fig. 3, some of our models show a final mass-loss rate that is so high ($\ge 10^{-4}M_{\odot}  \ \mathrm{yr}^{-1}$) that they may be observed as SNe IIn (\S 1). The observational properties of Type IIn SNe are known to be diverse, and the progenitors of this type of SNe are not well clarified \citep{2013A&A...555A..10T}. Then, it is possible that among the observed SNe IIn, there are some whose CSM are produced through this path, i.e. the mass loss before the explosion is driven by non-conservative mass transfer in the binary evolution.

Based on our binary models, let us estimate the expected rate of SNe IIn whose CSM is produced by non-conservative mass transfer. For simplicity, we fix the primary star to be $16M_{\odot}$, and assume the distribution of the initial periods as $f(P) \propto P^{-1}$ \citep{2007A&A...474...77K}.

The range of the initial period for which SNe IIb are produced is 10 days $\lesssim P \lesssim $ 1000 days (\S 3). Fig. 9 shows the mass-loss rates averaged over the final 1000 yr ($\dot{M}_{1000}$) as a function of the initial period. Assuming that the models with the final mass-loss rates of $\dot{M}_{1000}\gtrsim 10^{-4}M_{\odot}  \ \mathrm{yr}^{-1}$ explode as SNe IIn \citep{2013A&A...555A..10T,2014MNRAS...439..2917M} \footnote[3]{Assuming that the velocity associated with the mass loss is $\sim 10$ km s$^{-1}$, this corresponds to $\dot{M}_{1000} \gtrsim 10^{-3}M_{\odot}  \ \mathrm{yr}^{-1}$ if the velocity is assumed to be the frequently adopted value of $\sim 100$ km s$^{-1}$ for SNe IIn.}, a primary star in a binary system with initial period in the range 1800 days $\lesssim P \lesssim$ 2100 days would explode as an SN IIn. Therefore, the ratio of the number of the SNe IIn in the binary origin, whose CSM is produced by non-conservative mass transfer, to SNe IIb (from the binary path) is $(\mathrm{ln} 2100 - \mathrm{ln} 1800)/(\mathrm{ln} 1000 - \mathrm{ln} 10) \sim 0.033$.
The observed fraction of SNe IIb to all the CCSNe is 10.6\% \citep{2011MNRAS.412.1522S}. So, if we assume that all the SNe IIb are produced from binary evolution, the fraction of SNe IIn from binary evolution (whose surrounding CSM is produced by non-conservative mass transfer shortly before explosion) to all the CCSNe is estimated to be $\sim 0.35\%$. In other words, considering the relative frequency of SNe IIn in all CCSNe being 8.8 \% \citep{2011MNRAS.412.1522S}, we conclude that the SNe IIn from binary evolution could occupy $\sim 4$ \% of all observed SNe IIn. This could be higher if the distribution of the initial period, which is not strongly constrained especially for systems with a wide orbit, were flatter toward larger $P$. 

 Thus, we conclude that a small but non-negligible fraction of the observed SNe IIn is occupied by those whose surrounding CSM is produced by binary non-conservative mass transfer in the final phase of the progenitor evolution. While the expected rate is relatively small, it is highly interesting to identify these kinds of SNe IIn observationally. Since this is a solid prediction from the binary evolution model for SNe IIb, identifying such a population in SNe IIn is important to clarify not only the origin of SNe IIn but the role of the binary evolution toward SNe IIb. Indeed, these types of SNe IIn are expected to have two features which could be distinguishable from other scenarios, e.g., a luminous blue variable (LBV)-like progenitor. (1) As compared to the popular LBV-like progenitor scenario, the velocity associated with the mass loss will be smaller by a factor of a few (e.g. wind from the main sequence companion) or by more than an order of magnitude (e.g. from the giant progenitor). This is an interesting target for high-dispersion spectroscopy of nearby and bright SNe IIn \citep[or they may look like SNe IIL if the narrow absorption created within the CSM is contaminated by an unrelated background][]{2016MNRAS...456..323K}. (2) The CSM which is produced in this way may well have characteristic structure, because of its origin in the binary evolution, which has a specific axis as defined by the orbital plane. Uncovering the geometry of CSM around SNe IIn may allow us to confirm this idea \citep[see, e.g., ][]{2016ApJ...832..194K}.

 \begin{figure}[htbp]
   %%\hspace*{10mm}
  \centering
  \includegraphics[width=80mm]{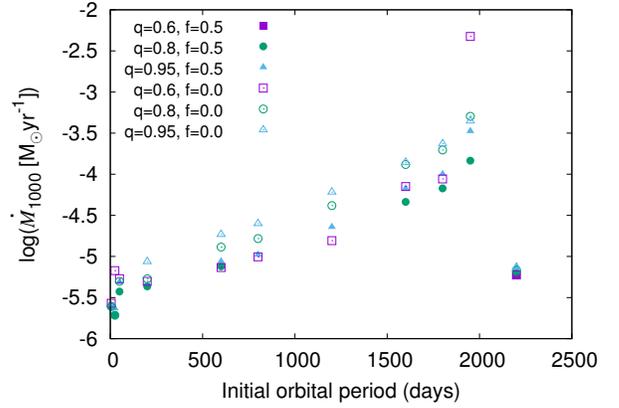}
  %%\vspace*{1.5cm}
  \caption{Average mass-loss rate in the final $1\times 10^3$ years as a function of the initial period.}
\end{figure}

\subsubsection{A population of SNe that interact with a shell-like CSM}

One of our models (No.28 in Table3) shows a very high mass-loss rate, showing as much as $\dot{M}_{1000} \sim 5 \times 10^{-3} M_{\odot} \ \mathrm{yr}^{-1}$. In this model, an intensive and unstable mass transfer begins very shortly before
the explosion. In general, when the primary begins to experience RLOF for the first time, the mass
transfer usually occurs on thermal or dynamical timescales, depending on whether their envelopes are radiative
or convective. Both timescales are usually much shorter than the evolutionary timescales. Shortly after this rapid mass transfer phase, stable mass transfer takes place (Section 4.1). In most of the models, this first intensive mass transfer occurs well before the collapse. Therefore, the material lost during this phase, if any, will have already gone too far to interact with the SN ejecta during the observable time window after the progenitor explodes as an SN. However, in some of the case C models, this intensive mass transfer occurs so shortly before the explosion ($\lesssim 10^3$ years), that we can observe the interaction of the SN ejecta with the CSM at the immediate vicinity of the progenitor as created by such an intensive mass loss. In this case, the shape of the CSM will probably be shell-like, unlike other models which have sustained mass loss for a long period before the explosion.

 In Fig. 10, the mass transfer rate evolution in the final $\sim 10^4$ years is shown for two representative models with case C mass transfer to illustrate this situation. The left and right panels show the models with $P = 1800$ days and $P = 1950$ days, respectively. The initial mass ratio and mass accretion efficiency are the same in both models. The model with $P = 1800$ days has a smaller Roche lobe radius, therefore the mass transfer sets in earlier. The model with $P = 1950$ days, which corresponds to No.28 in Table3, has a relatively larger Roche lobe radius, and significant mass transfer sets in only when the primary is about to explode; thus it may produce a shell-like CSM near the SN progenitor. Although the average mass-loss rate during the final $\sim 10^3$ yrs of this model is $\dot{M}\sim 5 \times 10^{-3} M_{\odot} \ \mathrm{yr}^{-1}$ (Table3), the temporary mass-loss rate can reach as much as $\dot{M}\gtrsim 10^{-2} M_{\odot} \ \mathrm{yr}^{-1}$ (Fig. 10). Thus, exploring the probability of such an unstable mass transfer taking place shortly ($\lesssim 10^3$ yr) before the collapse seems to be worthwhile.
\begin{figure}[htbp]
  \begin{minipage}{0.5\hsize}
%%\hspace*{10mm}
    \begin{center}
     \includegraphics[width=80mm]{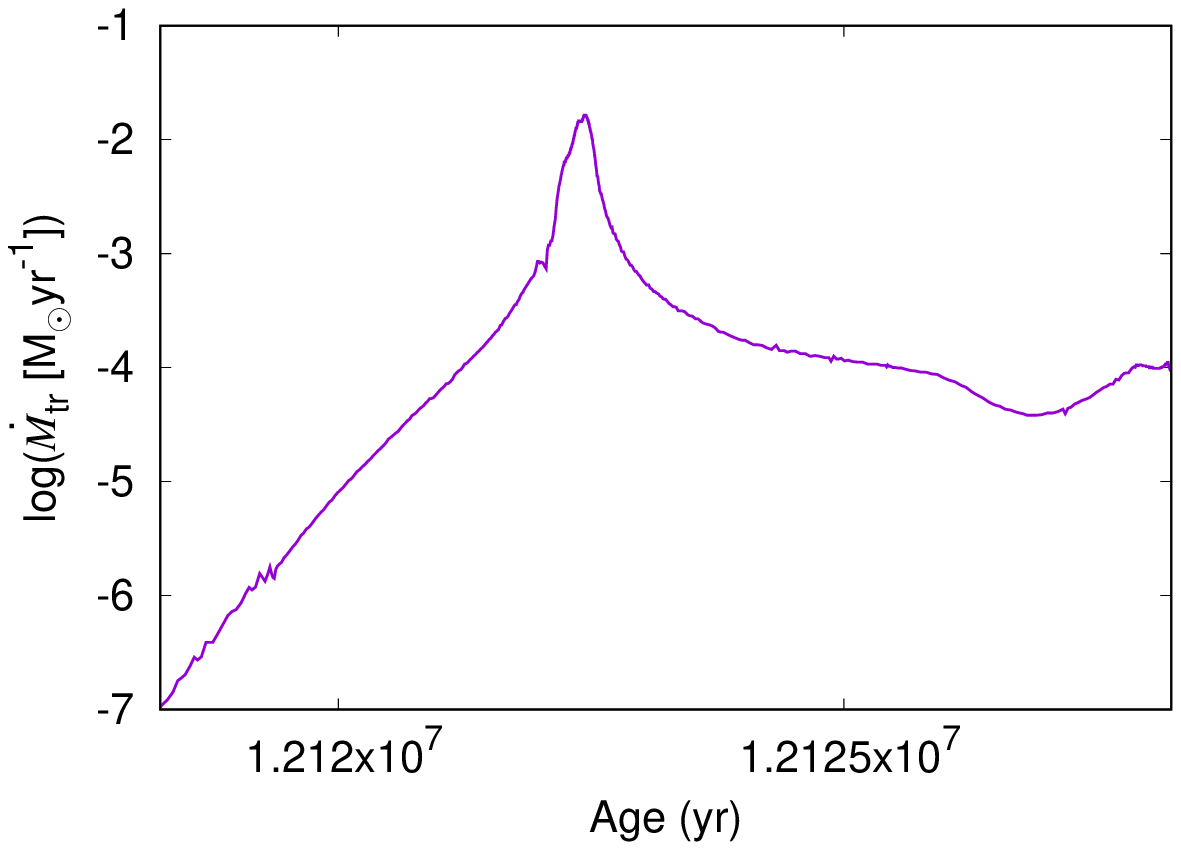}
    \end{center}
  \end{minipage}
  \begin{minipage}{0.5\hsize}
%%\hspace*{10mm}
    \begin{center}
      \includegraphics[width=80mm]{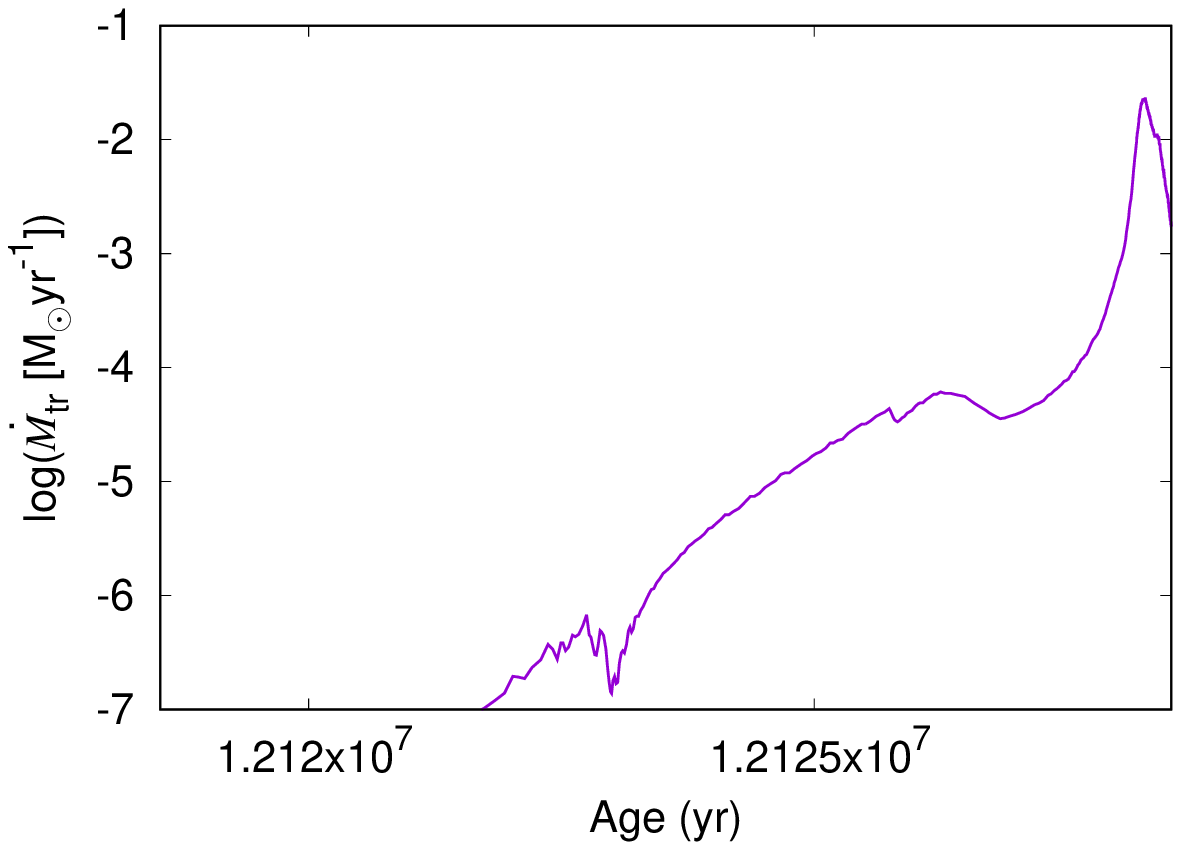}
    \end{center}
  \end{minipage}
  %%\vspace*{1.4cm} 
  \caption{Time evolution of the mass transfer rate ($M_{tr}$) for models No. 25 (left) and No, 28 (right) in Table 3 during the last 1 $\times 10^4$ yr.}
\end{figure}

Let us estimate the rate of such events. For the SN ejecta to interact with the shell-like CSM, intensive, unstable mass transfer should begin for the first time sometime during the last $\sim 10^3$ yr before the explosion. Fig. 11 shows the time evolution of the radius of a single $16M_{\odot}$ star in the last $5\times 10^3$ yrs of evolution, calculated under the same physical conditions as the primary stars in the binary models. In the last $\sim 10^3$ yrs, the radius changes from $905 R_{\odot}$ to $920 R_{\odot}$. In order for such an event to occur, the progenitor's Roche lobe radius need to be in this range. 

\begin{figure}[htbp]
  %%\hspace*{10mm}
  \begin{center}
  \includegraphics[width=80mm]{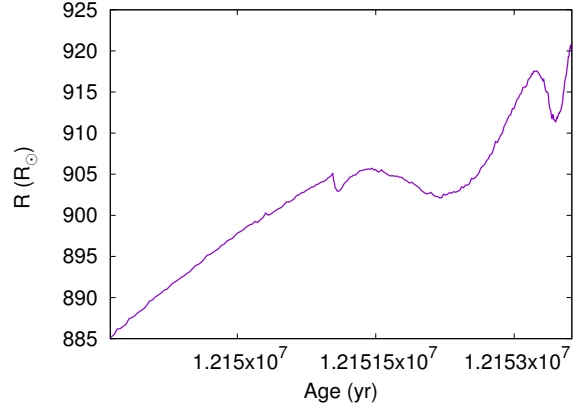}
  \end{center}
%%\vspace*{1.35cm}
  \caption{Time evolution of the radius of the $16M_{\odot}$ single-star model in the last $5\times 10^3$ yrs of the evolution.}
\end{figure}

In MESA, the Roche lobe radius of the primary is calculated as
\begin{eqnarray}
R_{\mathrm{rl}}=\frac{0.49q^{-\frac{2}{3}}}{0.6q^{-\frac{2}{3}}+\mathrm{ln}(1+q^{-\frac{1}{3}})} a \    .  
\end{eqnarray}
Here, $a$ is the binary separation expressed as
\begin{eqnarray}
a=\Bigl\{\frac{G(M_1+M_2)P^2}{4\pi^2} \Bigr\}^{\frac{1}{3}}  \   .
\end{eqnarray}
This can be calculated as
\begin{eqnarray}
\frac{R_{\mathrm{rl}}}{R_{\odot}}=F(q) (\frac{M}{16M_{\odot}})^{\frac{1}{3}} (\frac{P}{1day})^{\frac{2}{3}} \    .
\end{eqnarray}
Here, we define $F(q)$ as
\begin{eqnarray}
F(q)=\frac{5.19q^{-\frac{2}{3}}(1+q)^{\frac{1}{3}}}{0.6q^{-\frac{2}{3}}+\mathrm{ln}(1+q^{-\frac{1}{3}})}  \    .
\end{eqnarray}

For simplicity, we neglect the wind. In this case, until the beginning of intensive mass transfer, the mass of each star and orbital period remain constant. Assuming that mass transfer begins as soon as $R$ reaches $R_{\mathrm{rl}}$, the condition for the initial period leading to a dense CSM in the vicinity of the SN progenitor is described as follows: 

\begin{eqnarray}
F(q)^{-\frac{3}{2}}\times 905^{\frac{3}{2}} \lesssim P/\mathrm{day} \lesssim F(q)^{-\frac{3}{2}}\times 920^{\frac{3}{2}}  \  .
  \end{eqnarray}

The range of the initial period to satisfy this condition corresponds to $\Delta$ ln ($P$/day) = ln $(920)^{\frac{3}{2}}$ - ln $(905)^{\frac{3}{2}} \sim 0.025$. Note that $F(q)$ is canceled out. Comparing this with the corresponding value for Type IIb SNe, i.e., $\Delta$ ln ($P$/day) $\sim$ 4.6, the fraction of the SNe IIn which are likely to have very dense CSM produced by unstable mass transfer ($\dot{M} \gtrsim 10^{-2} M_{\odot} \mathrm{yr}^{-1}$) shortly before the explosion is $5 \times 10^{-3}$ times that of SNe IIb, or $\sim 0.06$ \% of all the observed CCSNe. This covers roughly 0.65 \% of all the observed SNe IIn.

The `volumetric' (intrinsic) rate as estimated above may sound like a prediction that would be impractical to confirm, perhaps marginally being testable only with large future surveys like the Large Synoptic Survey Telescope (LSST). However, we argue that this is not the case. The luminosity of SNe IIn, or in general SNe powered by SN-CSM interaction, is scaled to be roughly proportional to the CSM density \citep{2014MNRAS...439..2917M}. This is more complicated in the case of an optically thick, dense CSM \citep{2014ApJ...790L..16M}, while the argument should in any case apply to the total energy budget. The mass-loss rate of $\gtrsim 10^{-2} M_{\odot}$ yr$^{-1}$ in the binary systems discussed here (Fig. 10) corresponds to the CSM density by at least one order of magnitude larger than the less extreme SNe IIn with $10^{-4}$--$10^{-3} M_{\odot}$ yr$^{-1}$ (\S 4.2.1).

  Therefore, the luminosity of the SNe IIn under consideration (or SNe IIP/IIL which turn their appearance from SNe IIn to other types \citep[e.g.,][]{arXiv170102596}) could be extremely bright and occupy a signification fraction of `observed' SNe IIn in a magnitude-limited sense. Indeed, mass-loss rates exceeding $\sim 10^{-2} M_{\odot}$ yr$^{-1}$ have been derived for the most luminous SNe IIn \citep{2009Natur.458..865G, 2010MNRAS.404..305M}. Assuming that the luminosity of the SNe IIn in this class is larger than the less extreme case by two orders of magnitude (i.e., 5 mag), which is consistent with the luminosity function of SNe IIn in observed samples \citep{2011MNRAS...412..1441L}, then the detectable volume of such luminous SNe IIn in the universe is three orders
of magnitudes larger than that of the less extreme SNe IIn. Note that deriving the mass-loss rate may include an error at a level of an order of magnitude, but the diversity discussed here is well beyond such uncertainty. Therefore, despite the intrinsically rare occurrence of
these luminous SNe IIn, this population could indeed dominate, or at least significantly contribute
to, the observed SNe IIn from the binary evolution or even the observed SNe IIn as a whole.

%% Putting eqnarrays or equations inside the mathletters environment groups
%% the enclosed equations by letter. For instance, the eqnarray below, instead
%% of being numbered, say, (4) and (5), would be numbered (4a) and (4b).
%% LaTeX the paper and look at the output to see the results.

\section{Summary} \label{sec:summary}

Several progenitors of Type IIb SNe have been identified so far. Among these, four SNe have abundant observational data sets already published, including their location in the HR diagram and the mass-loss rates shortly before the explosion. In addition to the diversity in the HR diagram, there is a tendency that their mass-loss rates increase by an order of magnitude with the increase of the progenitor radii. In particular, the high mass-loss rates associated with the more extended progenitors are not readily explained by a prescription commonly used for a single stellar wind.

We have calculated a grid of binary evolution models with various parameter sets. We have shown that the observational relation between the progenitor radii and mass-loss rates can naturally be explained by non-conservative mass transfer in the final phase of progenitor evolution without any fine tuning.

 We have also clarified that the mass transfer rate in the final $\sim 10^3$ yr can be approximately estimated using an analytical formula (eq (6)), which roughly claims that the progenitor loses envelope mass within the timescale of the expansion of the radius. Using this formula, we can explain why the mass transfer rate increases with the progenitor radius (Fig. 3). This is mainly because less extended progenitors have not only a smaller envelope mass to transfer but a larger value of $\zeta_{\mathrm{eq}}$. The larger $\zeta_{eq}$ means that the progenitor shrinks faster in response to the mass loss.

This is further support for the dominance of binary evolution origin leading to Type IIb SNe. Therefore, further testing the relation between the size of the progenitor and the associated mass-loss rate with an increasing number of observed samples can provide a key to clarifying the still-debated origin toward SNe IIb, and eventually also to SNe Ib/c. 

As a byproduct, we have also found a possible link between the binary evolution scenario toward SNe IIb and some SNe IIn. About $4$ \% of all observed SNe IIn should have CSM which is produced by binary non-conservative mass transfer in the final evolutionary stage of the progenitor, if the main path to SNe IIb is the binary interaction. Such a population of SNe IIn will have characteristics of the velocity and geometry of the CSM, and therefore will be distinguishable from other SNe IIn from different evolutionary scenarios. Identifying such SNe IIn will provide a new test for the binary origin toward SNe IIb (and a fraction of SNe IIn, or even SNe IIL).

 Furthermore, about one tenth of SNe IIn related to binary evolution are predicted to be associated with an extensively dense mass loss, reaching $\dot{M} \gtrsim 10^{-2} M_{\odot} \mathrm{yr}^{-1}$ in the final $\sim 10^3$ yr, which likely produces a shell-like CSM in the immediate vicinity of the SN progenitor. They may indeed be classified as either SNe IIP or IIL, but initially showing the spectroscopic features of SNe IIn. While the intrinsic (volume-limited) rate is predicted to be small, these may dominate the observed (magnitude-limited) sample of SNe IIn through the binary path (for which the CSM is produced by binary non-conservative mass transfer), or even a large fraction of all the luminous SNe IIn.

\acknowledgments
The authors thank Sung-Chul Yoon, Takashi Moriya, Ryosuke Hirai, Akihiro Suzuki and Takashi Nagao for useful comments on this research and stimulating discussions, and also thank the anonymous referee for the constructive comments on this manuscript. The authors thank the Yukawa Institute for Theoretical Physics at Kyoto University for useful discussions to complete this work during the YITP workshop YITP-T-16-05 on `Transient Universe in the Big Survey Era: Understanding the Nature of Astrophysical Explosive Phenomena'. The work of K. M. has been supported by Japan Society for the Promotion of Science (JSPS) KAKENHI Grant 26800100 and 17H02864 (K.M.).
\appendix

\section{The relation between the hydrogen-rich envelope mass and the radius in the equilibrium state}

The radius decreases with the decrease of the envelope mass when the envelope is radiative (ln ($M_{\mathrm{env}} [M_{\odot}]$) $\lesssim$ -2) as shown in Fig. 6. In this Appendix, we show that this behavior can be described approximately in an analytical way, following an argument similar to that presented by \citet{1961ApJ...133..764C} in a different context. 
First, the basic equations determining the structure of the stellar envelope are 
\begin{eqnarray}
\frac{d (P_{\mathrm{gas}}+P_{\mathrm{rad}})}{d r} &=& - \frac{G M}{r^2} \rho \ ,   \\
\frac{d P_{\mathrm{rad}}}{dr} &=& -\frac{\kappa L}{4 \pi c r^2} \rho \ .
\end{eqnarray}
Here, $P_{\mathrm{gas}}$ and $P_{\mathrm{rad}}$ are the pressure of the gas and radiation, respectively. For the opacity ($\kappa$), we assume that free-free absorption is the dominant source of the opacity, therefore $\kappa=\kappa_0 \rho T^{-7/2}$. Here, we also assume that the mass and energy generation in the envelope are negligible compared with those in the core and the surrounding shell. We further assume for simplicity that $P_{\mathrm{gas}} = \beta P$, with $\beta$ being constant throughout the envelope. Using these relations, we can solve Equations (A1) and (A2) analytically: 
\begin{eqnarray}
  \rho &=& C  \left(\frac{1}{r}- \frac{1}{R}\right)^{13/4} \ , {\rm and}\\
  C &=& \sqrt{\frac{16 \pi a c}{3 \kappa_0 L}} \left(\frac{4 G M \mu H \beta}{17 k_{\mathrm{B}}} \right)^{15/4} \ ,
\end{eqnarray}
where $\mu$ is the mean molecular weight, while $k_{\mathrm{B}}$ and $H$ are Boltzmann's constant and the reciprocal of Avogadro's number, respectively. From this solution, we can express the hydrogen-rich envelope mass as follows: 
\begin{eqnarray}
  M_{\mathrm{env}} &=& \int_{R_{\mathrm{c}}}^{R} dr \ 4 \pi r^2 \rho \ ,  \\
  M_{\mathrm{env}}/M_{\odot} &=& 4 \pi C \ ' I(z) \ , 
\end{eqnarray}
where 
\begin{eqnarray}
  C \ ' &=& \frac{4 \pi C}{M_{\odot} R_{\mathrm{c}}^{1/4}} \ , \\
  I(z)&=& \frac{z^{17/4}}{(1+z)^{1/4}} \int_0^1 du \ \frac{u^{13/4}}{(1+zu)^4} \ .
\end{eqnarray}
$R_{\mathrm{c}}$ and $R$ are the radius of the helium core and the stellar radius, respectively. Here, we define a new variable $z=\frac{R-R_{\mathrm{c}}}{R_{\mathrm{c}}}$. From (A6)--(A8), and substituting the typical values of the models (Table 4) for the corresponding physical parameters in (A4) and (A7), we obtain approximately the radius as a function of the envelope mass. In Fig. 12, we compare this analytically derived relation with the equilibrium radius derived in Section 4. 1 (Fig. 6). Despite the crude approximations (the opacity is dominated by the free-free absorption and the ratio of the gas pressure to the radiation pressure is constant throughout the envelope), the analytical curve derived from (A6) reproduces the curve obtained through the numerical evolution calculations fairly well. 
Thus, the radius in complete equilibrium increases with increasing envelope mass when the envelope is radiative under the conditions we assumed, which is then suppressed due to the development of convection for ln ($M_{\mathrm{env}} [M_{\odot}]$) $\gtrsim$ -2.

\section{Asymptotic behavior of $\zeta_{\mathrm{L}}$}

For a large mass ratio ($q \gg 1$), which is often expected in the final stage of the binary evolution in the situation we are considering, $\zeta_{\mathrm{L}}$ converges to $\sim -1.5$. This is explained as follows. Instead of using the fitting formula of \citet{1983ApJ...268..368E}, here our argument is based on the Roche potential. If we set the origin of the coordinates at the center of the primary (i.e. star 1), and set the y-axis and z-axis parallel to the line connecting the two stars and the orbital rotation axis, respectively, the Roche potential $\phi$ is written as
\begin{eqnarray}
  \phi (x, y, z) = -\frac{GM_1}{(x^2+y^2+z^2)^{1/2}} -\frac{GM_2}{((x-a)^2+y^2+z^2)^{1/2}}-\frac{1}{2}\Omega^2 [(x-\mu a)^2+y^2] \  .
\end{eqnarray}
$\mu$ is defined as $\mu = \frac{M_2}{M_1+M_2}$, and $a$ and $\Omega=\sqrt{\frac{G(M_1+M_2)}{a^3}}$ are the binary separation and the angular velocity of the orbit, respectively. If we denote the position of the L1 point as ($x_{L1}$,0 ,0), then $x_{L1}$ is derived from
\begin{eqnarray}
  \frac{\partial \phi (x, 0, 0)}{\partial x} =0 .
\end{eqnarray}
Noting that $0<x_{L1}<a$, this leads to
\begin{eqnarray}
  \frac{G M_1}{x^2}-\frac{G M_2}{(x-a)^2} - \Omega^2 (x-\mu a)=0 .
\end{eqnarray}
If $q$ is sufficiently large, then we expect $x_{L1} \ll a$. In this limit, Equation (19) is expanded in terms of $(x/a)$ as follows: 
\begin{eqnarray}
  1 &=& (1+3q) \left(\frac{x}{a} \right)^3 + O\left(\left(\frac{x}{a}\right)^4\right)  \\
  &\sim& 3q \left(\frac{x}{a}\right)^3 .
\end{eqnarray}
Thus, we obtain 
\begin{eqnarray}
  \frac{x_{L1}}{a} \propto q^{-\frac{1}{3}} .
\end{eqnarray}
Denoting the orbital angular momentum as $J$, $a$ is expressed as
\begin{eqnarray}
  a &=& \frac{J^2}{G} \frac{M_1+M_2}{M_1^2M_2^2} \\
  &=& \frac{J^2}{G} \frac{M_2}{M_1^2}    \  (q \gg 1) \ .
\end{eqnarray}

When $q$ is sufficiently large, $M_2$ is practically constant, even with mass transfer from the primary. Moreover, because we assume that the specific angular momentum of the escaping material has a value identical to that of the accreting star (secondary) in our calculations, the loss of angular momentum is negligible if $q \gg 1$. Then, from equations (B14) and (B16), we obtain the following: 
\begin{eqnarray}
  R_{rl, 1} &\sim& x_{L1}   \\
  &\propto& M_1^{-5/3}
  \end{eqnarray}
Therefore, we finally reproduce the asymptotic behavior found in the binary evolution calculations: 
\begin{eqnarray}
\zeta_{\mathrm{L}} \sim -\frac{5}{3} \ (q \gg 1) .
\end{eqnarray}
Thus, the absolute value of the $\zeta_{\mathrm{L}}$ does not exceed 5/3, and typically has the value of $\zeta \approx -1$ for large values of $q$ relevant to the situation considered in this paper.

\begin{figure}[htbp]
  %%\hspace*{10mm}
  \centering
  \includegraphics[]{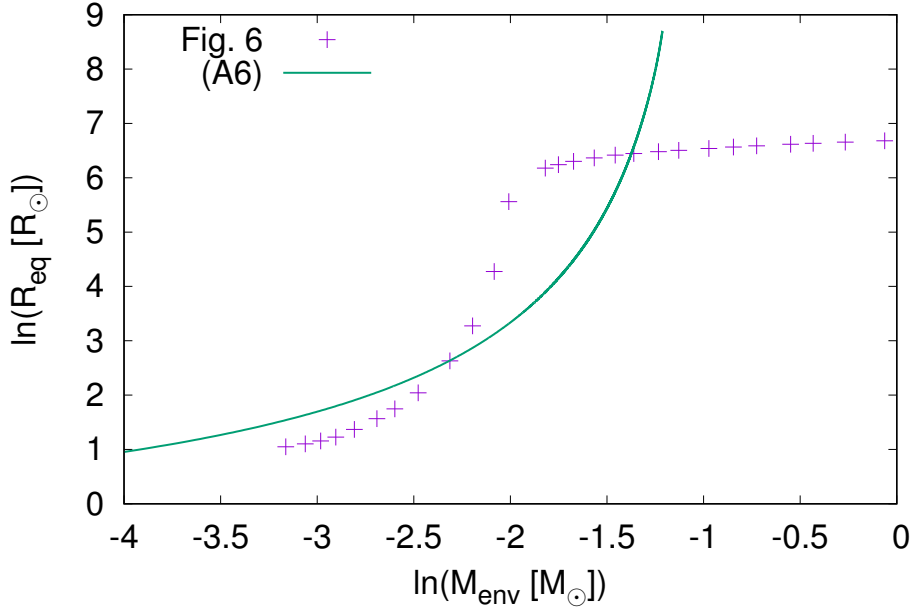}
  %%\vspace*{1.5cm}
  \caption{Relation between the stellar radius and the hydrogen-rich envelope mass under dynamical and thermal equilibrium, for a particular model sequence considered in this paper as derived numerically through the evolution calculations (points from Fig. 6). Also shown is the relation derived analytically in this section assuming a radiative envelope, for the parameters corresponding to the numerical models (line). Note that the discrepancy between the two treatments becomes larger as convection is developed in the envelope.}
\end{figure}

\begin{table}[htbp]
  \centering
  \caption{Final properties of the models ($f=0.5$). The $M_{\rm env}$ and $M$ denote the hydrogen-rich envelope mass and star mass, respectively. $\dot{M}_{\mathrm{wind}}$ is the mass-loss rate due to the stellar wind at the end of the calculations. The subscript 1 and 2 refer to the primary and the secondary, respectively, while the subscript f refers to the values at the end of the calculation. 'NON CONVERGENCE' in the column of the final fate means the calculation has a convergence problem during the calclation.}
  \vspace{0.75cm}
  \hspace*{-4cm}
  \scalebox{0.7}{
    \begin{tabular}{|l|c|c|c|c|c|c|c|c|c|c|r|} \hline 
      No. & Initial period (P) & Initial mass ratio (q)  & log$T_{\mathrm{eff}, 1}$ & log$(\frac{L_{1}}{L_{\odot}})$ &  Radius $R_1$     &  $M_{\mathrm{env}, 1}$    & $M_1$ & $M_2$  & $\dot{M}_{\mathrm{wind}} $          &　$\dot{M}_{1000}$  & Final fate  \\ 
      &  (Days)  &     &             &                 &  $(R_{\odot})$ &  $(M_{\odot})$  & $(M_{\odot})$ & & $(10^{-6}M_{\odot} \mathrm{yr}^{-1})$ & $(10^{-6}M_{\odot} \mathrm{yr}^{-1})$ & of the primary   \\ \hline \hline
      1 &  5  & 0.6  & -     & -     & -    & -      & -       & -    & - & -  & CONTACT   \\ \hline
      2 &  5  & 0.8  & 4.53  & 4.85  & 7.6  &  0.000 & 3.801 & 17.804  &1.99  & 2.43 & SN Ib \\ \hline
      3 & 5   & 0.95 & 4.54  & 4.85  & 7.5  & 0.000  & 3.816 & 19.937  &2.07  & 2.51 & SN Ib \\ \hline
      4 & 25  & 0.6  & -     & -    & -      &  -      &  -  & -   & -   & -   & CONTACT \\ \hline
      5 & 25   & 0.8 & 4.24  & 4.94  & 32.11 & 0.059  & 4.533 & 17.864  & 1.57 & 1.93  & SN IIb  \\ \hline
      6 & 25   & 0.95 &  4.43 &  4.94  & 13.4  &  0.043 & 4.530 & 19.857 & 1.94  &2.33 & SN IIb  \\ \hline
      7 &  50  & 0.6 &  -     & -     & -       & -      & -     & - & -  & -   & CONTACT \\ \hline
      8 &  50  & 0.8 &   3.90 & 4.94  & 156.3   & 0.057  & 4.562 & 18.005  & 1.68  & 3.74 & SN IIb  \\ \hline
      9 &  50  & 0.95  & 3.87  & 4.95  & 178.5  & 0.057 & 4.607  & 20.030 &2.53  & 4.88 & SN IIb \\ \hline
      10 &  200  & 0.6 & -     & -      & -       & -     & -   & -   & -   & -   & CONTACT \\ \hline
      11 &  200  & 0.8 & 3.72 &4.95    & 364.0  & 0.065  &  4.611 & 17.957  & 2.50  & 4.32 & SN IIb \\ \hline
      12  &  200  & 0.95 & - & - & -   &-  &  -  &- & - & - & NON CONVERGENCE \\ \hline
      13  &  600  & 0.6  & -    & -     & -        & -      & -   & -    & -   & -    & CONTACT \\ \hline
      14  &  600  & 0.8  &   3.61   &  4.99  & 626.4  & 0.151 & 4.962 & 17.585  & 3.18 & 7.58  & SN IIb \\ \hline
      15  &  600  & 0.95  & 3.60   &  5.00 & 649.3 & 0.168  & 5.048  & 19.376  &  3.36 & 8.72  &  SN IIb \\ \hline
      16  &  800  & 0.6   & -     & -     & -      & -      &  -  & -  &  -  & -   & CONTACT   \\ \hline
      17  &  800  & 0.8   &  -    & -     & -      &  -     & -  & -    & -   & -   & CONTACT \\ \hline
      18  &  800  & 0.95  & 3.59  & 5.00  & 688.4  &  0.212 & 5.156 & 18.942 &  3.61 &  10.51    &  SN IIb  \\ \hline
      19  &  1200 & 0.6   & -     & -     & -      & -      & -   & -   & -   & -    & CONTACT \\ \hline
      20  &  1200 & 0.8   & -     & -     &  -     &  -     & -  & -    & -   & -    & CONTACT \\ \hline
      21  &  1200 & 0.95  & 3.55 &  5.01  & 843.4  &  0.714 & 5.746 &  18.369 &  4.94  & 22.83 & SN IIb   \\ \hline
      22  &  1600 & 0.6   & -     & -     & -      & -      & -  & -   & -   & -    & CONTACT \\ \hline
      23  &  1600 & 0.8   &  3.53   &   5.02   & 948.2   &  2.040    &  7.127 & 15.488  &  5.53  &  46.10  &SN IIP/IIL  \\ \hline
      24  &  1600 & 0.95  & 3.53  & 5.02 & 960.0  & 2.768  & 7.875 & 16.752  & 6.99 & 67.13 & SN IIP/IIL \\ \hline
      25  &  1800 & 0.6   &  -  & -   &  -  &  -   &  -    & - & -  & - &  CONTACT   \\ \hline
      26  &  1800 & 0.8   &  3.52  &  5.020  & 965.07   &   2.714   & 7.802 & 15.145 & 5.58 & 67.27 & SN IIP/IIL \\ \hline
      27 &   1800 & 0.95 & 3.53 & 5.02 & 955.9 & 3.725 & 8.832  & 16.274 &  6.86 & 100.30 & SN IIn  \\ \hline
      28  &  1950 & 0.6   & -   &  -  &  -  &  -   &  -    &  -  &  - & - & CONTACT \\ \hline
      29  &  1950 & 0.8   &  3.52  & 5.02  &   965.2 &  3.408  & 8.498   & 14.799 & 5.65  & 146.32 & SN IIn \\ \hline
      30 &   1950 & 0.95 & 3.53 & 5.02 & 936.7 &  5.030 &  10.14 & 15.625 & 6.90 & 334.51 & SN IIn  \\ \hline
      31  &  2200 & 0.6 & 3.54 & 5.02 & 890.3 & 7.553  &12.664 & 9.548  & 5.09 &  5.94 & SN IIP/IIL  \\ \hline
      32  &  2200 & 0.8 & 3.54  & 5.02 & 887.9 & 7.669  & 12.757 & 12.670  & 5.30 & 6.40 & SN IIP/IIL \\ \hline
      33  &  2200 & 0.95 &   3.54 & 5.02 & 890.8  & 7.549 & 12.657 & 14.385 & 6.52 & 7.30 & SN IIP/IIL \\ \hline
    \end{tabular}
 }
\end{table}

\begin{table*}[htbp]
  \centering
  \caption{Final properties of the models ($f=0.0$). The term `SECONDARY'S RLOF' in the column of the final fate means that the secondary initiated mass transfer after the mass transfer by the primary finished but before the core collapse.}
  \vspace{0.75cm}
   \hspace*{-4cm}
  \scalebox{0.7}{
    \begin{tabular}{|l|c|c|c|c|c|c|c|c|c|c|r|} \hline
      No. & Initial period (P) & Initial mass ratio (q)  & log$T_{\mathrm{eff}, 1}$ & log$(\frac{L_{1}}{L_{\odot}})$ &  Radius $R_1$     &  $M_{\mathrm{env}, 1}$    & $M_1$ & $M_2$  & $\dot{M}_{\mathrm{wind}} $          &　$\dot{M}_{1000}$  & Final fate   \\ 
      & (Days)  &     &             &                 &  $(R_{\odot})$ &  $(M_{\odot})$  & $(M_{\odot})$ & & $(10^{-6}M_{\odot} \mathrm{yr}^{-1})$ & $(10^{-6}M_{\odot} \mathrm{yr}^{-1})$ &  of the primary \\ \hline \hline
      1 &  5  & 0.6  & 4.58   &  4.89    &  6.4  &  0.000    &   4.084  &  9.548  & 2.20    & 2.70   & SN Ib   \\ \hline
      2 &  5  & 0.8  &  4.54  &  4.85 & 7.4 &  0.000 & 3.832  & 12.659 & 2.03 & 2.48 & SN IIb \\ \hline
      3 & 5   & 0.95 &  - & - & -  & -  & - &-  &-  & - & SECONDARY'S RLOF \\ \hline
      4 & 25  & 0.6  & 4.07   & 4.95   & 71.2     &  0.057    & 4.587 & 9.548 & 1.42   & 6.72  & SN IIb \\ \hline
      5 & 25   & 0.8 & 4.21  & 4.94  & 37.0 & 0.062  & 4.538  & 12.670 & 1.58 & 1.92 &  SN IIb \\ \hline
      6 & 25   & 0.95 & -  & -  & - & - & - & - &- & - & SECONDARY'S RLOF   \\ \hline
      7 &  50  & 0.6 &  3.97    &  4.94   &  111.3     &  0.055    &  4.581 & 9.548  &  0.66  & 5.36   & SN IIb \\ \hline
      8 &  50  & 0.8 &  3.88   & 4.94  & 168.4  & 0.056 & 4.565 & 12.670  & 0.99 &  4.98 &  SN IIb \\ \hline
      9 &  50  & 0.95  & -  & -  & - & - & - &-  &-  &  - &   SECONDARY'S RLOF \\ \hline
      10 &  200  & 0.6 &   3.79   &  4.95    &  261.9  &  0.058   & 4.602  & 9.548  & 1.38   &  4.99  & SN IIb \\ \hline
      11 &  200  & 0.8 & 3.7009 & 4.9481   & 394.0 & 0.068  & 4.620  & 12.670 & 2.00 & 5.35  & SN IIb \\ \hline
      12  &  200  & 0.95 & 3.66 & 4.95 &  474.6  &  0.084  & 4.676  & 14.326 & 4.41 & 8.65 &  SN IIb \\ \hline
      13  &  600  & 0.6  &  3.65   & 4.96     &  507.0    &   0.092   & 4.714   & 9.548 & 2.49  & 7.32   & SN IIb \\ \hline
      14  &  600  & 0.8  & 3.60  & 4.99 & 648.3  & 0.167 & 5.004  & 12.670 & 3.35 & 12.99  & SN IIb \\ \hline
      15  &  600  & 0.95  &  3.59  & 5.00  & 698.9  &  0.231 & 5.152 &  14.396 & 4.83 &  18.49 & SN IIb  \\ \hline
      16  &  800  & 0.6   &  3.62    &  4.98  &  580.6   &  0.123   &   4.852   & 9.548 &  2.86  & 9.88   & SN IIb   \\ \hline
      17  &  800  & 0.8   & 3.59   &  5.00    &  695.4    &   0.222   &  5.125   & 12.670  &  3.57  & 16.50  & SN IIb \\ \hline
      18  &  800  & 0.95  & 3.57  &  5.01 & 762.7  & 0.361  & 5.333 & 14.343 & 5.70 &  25.09    & SN IIb    \\ \hline
      19  &  1200 & 0.6   & 3.60  &  5.00    & 674.7    & 0.198    & 5.076 & 9.548 & 3.44    & 15.57    & SN IIb \\ \hline
      20  &  1200 & 0.8   &  3.55   & 5.01  &   856.2   &  0.777  &  5.769  & 12.670 & 4.89  &  41.51   & SN IIb \\ \hline
      21  &  1200 & 0.95  &  3.54 & 5.02  & 913.6   & 1.402  & 6.473 & 14.337 & 7.04 & 60.74 & SN IIP/IIL   \\ \hline
      22  &  1600 & 0.6   &  3.54    & 5.02  & 919.0   &  1.253   & 6.363 & 9.548 & 5.39  & 70.77    & SN IIP/IIL \\ \hline
      23  &  1600 & 0.8   &  3.53   & 5.02  &  958.8    &   2.851   &  7.937  & 12.670  &  5.63  & 131.45    & SN IIn \\ \hline
      24  &  1600 & 0.95  & 3.53  & 5.02 & 951.3  & 3.712  &  8.819 & 14.361 & 6.90 & 139.57 & SN IIn  \\ \hline
      25  &  1800 & 0.6   & 3.53   &  5.02  & 956.7   &  1.725   &   6.835  & 9.548 & 5.49   &  87.61  & SN IIP/IIL \\ \hline
      26  &  1800 & 0.8   &  3.53  &   5.02  & 948.9   &  3.879  &  8.968  & 12.669 & 5.51  & 197.10 & SN IIn  \\ \hline
      27 &   1800 & 0.95 & 3.53 & 5.01 & 924.8 & 4.945 &  10.053 & 14.363 & 6.60 & 234.26 & SN IIn  \\ \hline
      28  &  1950 & 0.6   &  3.52  &  5.02  &  969.7  &  2.725   &  7.836  & 9.548 &  5.52  &  4739.21  & SN IIn \\ \hline
      29  &  1950 & 0.8   & 3.53   &  5.02  &  931.9 &  5.179   & 10.267  & 12.669 & 5.49  & 506.14  & SN IIn \\ \hline
      30 &   1950 & 0.95 & 3.54 & 5.02 & 912.4 & 6.386 &  11.493 & 14.359 & 6.86 & 443.59 & SN IIn  \\ \hline
      31  &  2200 & 0.6 &  3.54 & 5.02 & 890.4 & 7.553  & 12.664 & 9.548 & 5.17 & 5.95 &  SN IIP/IIL \\ \hline
      32  &  2200 & 0.8 & 3.54 & 5.02 & 888.7  &  7.669 & 12.757 & 12.670 & 5.31 & 6.81  & SN IIP/IIL  \\ \hline
      33  &  2200 & 0.95 & 3.54 & 5.02 & 891.22  & 7.549 & 12.657 & 14.385 & 6.53 & 7.39  & SN IIP/IIL \\ \hline
    \end{tabular}
  }
\end{table*}

\begin{table}[htb]
  \centering
  \caption{Typical values of the physical quantities used in Fig. 12.}
  \begin{tabular}{|l|c|c|c|c|c|r|}  \hline
    $M/M_{\odot}$ & $L/L_{\odot}$ & $R_{\mathrm{c}}/R_{\odot}$ & $\beta$ & $\mu$ & $\kappa_0$ ($\mathrm{cm}^2 \mathrm{g}^{-1}$)  \\ \hline
   4.6 & $1.0\times 10^5 $ & 0.60 & 0.40  & 1.0 & $1.5 \times 10^{25}$ \\ \hline
  \end{tabular}
\end{table}


\begin{thebibliography}{999}
\bibitem [Benvenuto et al.(2013)]{2013ApJ...762...74B} {Benvenuto}, O.~G., {Bersten}, M.~C., {Nomoto}, K. 2013,  \apj, 762, 74
\bibitem [Chevalier \& Soderberg(2010)]{2010ApJ...711L..40C} {Chevalier}, R.~A. and {Soderberg}, A.~M. 2010, \apj, 711, 40
\bibitem [Cox \& Salpeter (1961)]{1961ApJ...133..764C} {Cox}, J.~P. and {Salpeter}, E.~E. 1961, \apj, 133, 764
  \bibitem [de Jager et al.(1988)]{1988A&AS...72..259D}  {de Jager}, C. and {Nieuwenhuijzen}, H. and {van der Hucht}, K.~A. 1988, A\&AS, 72, 259
\bibitem [Eggleton(1983)]{1983ApJ...268..368E} {Eggleton}, P.~P. 1983, \apj, 268, 368
  \bibitem [Filippenko(1997)]{1997ARA&A..35..309F} {Filippenko}, A.~V. 1997, ARA\&A, 35, 309
\bibitem [Folatelli et al.(2014)]{2014ApJ...793L..22F}  {Folatelli}, G. and {Bersten}, M.~C. and {Benvenuto}, O.~G. et al. 2014, \apj, 793, 22
  \bibitem [Folatelli et al.(2015)]{2015ApJ...811..147F} {Folatelli}, G., {Bersten}, M.~C., {Kuncarayakti}, H. et al. 2015, \apj, 811, 147
\bibitem [Fransson et al.(1996)]{1996ApJ...461..993F} {Fransson}, C. and {Lundqvist}, P. and {Chevalier}, R.~A. 1996, \apj, 461, 993 
\bibitem [Gal-Yam \& Leonard(2009)]{2009Natur.458..865G}  {Gal-Yam}, A. and {Leonard}, D.~C. 2009, Natur, 458, 865
\bibitem [Georgy(2012)]{2012A&A...538L...8G} {Georgy}, C. 2012, A\&A, 538, 8
\bibitem [Gr{\"a}fener \& Vink(2016)]{2016MNRAS.455..112G}  {Gr{\"a}fener}, G. and {Vink}, J.~S. 2016, MNRAS, 455, 112
\bibitem [Henyey et al.(1965)]{1965ApJ...142..841H} {Henyey}, L., {Vardya}, M.~S., {Bodenheimer}, P. 1965, \apj, 142, 841
 \bibitem [Herwig(2000)]{2000A&A...360..952H} {Herwig}, F. 2000, A\&A, 360, 952
 \bibitem [Ivanova(2015)]{2015ebss.book..179I} {Ivanova}, N. 2015, Ecology of Blue Straggler Stars, ed. H. M. J. Boffin, G. Carraro \& G. Beccari (Astrophysics and Space Science Library, Vol. 413; Berlin: Springer), 179
 \bibitem [Kamble et al.(2016)]{2016ApJ...818..111K} {Kamble}, A., {Margutti}, R., {Soderberg}, A.~M. et al. 2016, \apj, 818, 111
\bibitem [Kangas et al.(2016)]{2016MNRAS...456..323K} {Kangas}, T., Mattila, S., Kankare, E., et al. 2016, MNRAS, 456, 323 
\bibitem [Katsuda et al.(2016)]{2016ApJ...832..194K} {Katsuda}, S. and {Maeda}, K. and {Bamba}, A., et al. 2016, \apj, 832, 194
\bibitem [Kiewe et al.(2012)]{2012ApJ...744...10K} {Kiewe}, M. and {Gal-Yam}, A. and {Arcavi}, I., et al. 2012, \apj, 744, 10
\bibitem [Kolb \& Ritter(1990)]{1990A&A...236..385K} {Kolb}, U. and {Ritter}, H. 1990, A\&A, 236, 385
   \bibitem [Kouwenhoven et al.(2007)]{2007A&A...474...77K} {Kouwenhoven}, M.~B.~N. and {Brown}, A.~G.~A. and {Portegies Zwart}, S.~F., et al. 2007, A\&A, 474, 77
 \bibitem [Langer et al.(1985)]{1985A&A...145..179L} {Langer}, N., {El Eid}, M.~F., {Fricke}, K.~J. 1985, A\&A, 145, 179
\bibitem [Langer et al.(2003)]{2003astro.ph..2232L}  {Langer}, N. and {Yoon}, S.~-. and {Petrovic}, J., Heger A. 2003, in Maeder A., EenensP., eds, IAUSymp. 215, Stellar Rotation. Astron. Soc. Pac., SanFrancisco, in press (astro-ph/0302232) 
\bibitem [Li et al.(2011)]{2011MNRAS...412..1441L} {Li}, W. and {Leaman}, J. and {Chornock}, R., et al. 2011, MNRAS, 142, 1441
 \bibitem [Maeda et al.(2014)]{2014ApJ...785...95M} {Maeda}, K., {Katsuda}, S., {Bamba}, A., et al. 2014, \apj, 785, 95
 \bibitem [Maeda et al.(2015)]{2015ApJ...807...35M} {Maeda}, K., {Hattori}, T., {Milisavljevic}, D., et al. 2015, \apj, 807, 35
 \bibitem [Maund et al.(2004)]{2004Natur.427..129M} {Maund}, J.~R., {Smartt}, S.~J., {Kudritzki}, R.~P., et al. 2004, Natur, 427, 129
\bibitem [Maund et al.(2011)]{2011ApJ...739L..37M} {Maund}, J.~R., {Fraser}, M., {Ergon}, M., et al. 2011, \apj, 739, 37
\bibitem [Miller et al.(2010)]{2010MNRAS.404..305M}  {Miller}, A.~A. and {Silverman}, J.~M. and {Butler}, N.~R., et al. 2010, MNRAS, 404, 305
\bibitem [Moriya \& Maeda(2014)]{2014ApJ...790L..16M} {Moriya}, T.~J., {Maeda}, K. 2014, \apj, 790, 16
\bibitem [Moriya et al.(2014)]{2014MNRAS...439..2917M}  {Moriya}, T.~J., {Maeda}, K., Taddia, F., et al. 2014, MNRAS, 439, 2917
  \bibitem [Nugis \& Lamers(2000)]{2000A&A...360..227N}  {Nugis}, T. and {Lamers}, H.~J.~G.~L.~M. 2000, A\&A, 360, 227
\bibitem [Paxton et al.(2011)]{2011ApJS..192....3P} {Paxton}, B., {Bildsten}, L., {Dotter}, A., et al. 2011, ApJS, 192, 3
\bibitem [Paxton et al.(2013)]{2013ApJS..208....4P}{Paxton}, B., {Cantiello}, M., {Arras}, P., et al. 2013, ApJS, 208, 4
\bibitem [Paxton et al.(2015)]{2015ApJS..220...15P}{Paxton}, B., {Marchant}, P., {Schwab}, J., et al. 2015, ApJS, 220, 15
\bibitem [Petrovic et al.(2005)]{2005A&A...435.1013P} {Petrovic}, J. and {Langer}, N. and {van der Hucht}, K.~A. 2005, A\&A, 435, 1013
\bibitem [Sana et al.(2012)]{2012Sci...337..444S}{Sana}, H., {de Mink}, S.~E., {de Koter}, A. 2012, Sci, 337, 444
\bibitem [Smith et al.(2011)]{2011MNRAS.412.1522S} {Smith}, N., {Li}, W., {Filippenko}, A.~V., {Chornock}, R. 2011, MNRAS, 412, 1522
\bibitem [Soberman et al.(1997)]{1997A&A...327..620S} {Soberman}, G.~E., {Phinney}, E.~S., {van den Heuvel}, E.~P.~J. 1997A\&A, 327, 620
\bibitem [Stancliffe \& Eldridge(2009)]{2009MNRAS.396.1699S} {Stancliffe}, R.~J., {Eldridge}, J.~J. 2009, MNRAS, 396, 1699
\bibitem [Taddia et al.(2013)]{2013A&A...555A..10T} {Taddia}, F. and {Stritzinger}, M.~D. and {Sollerman}, J., et al. 2013, A\&A, 555A, 10
\bibitem [Van Dyk et al.(2014)]{2014AJ....147...37V} {Van Dyk}, S.~D., {Zheng}, W., {Fox}, O.~D. 2014, AJ, 147, 37
\bibitem [van Loon et al.(2005)]{2005A&A...438..273V}  {van Loon}, J.~T. and {Cioni}, M.-R.~L. and {Zijlstra}, A.~A. and 
  {Loup}, C. 2005, A\&A, 438, 273
  \bibitem [van Rensbergen et al.(2011)]{2011A&A...528A..16V}  {van Rensbergen}, W. and {de Greve}, J.~P. and {Mennekens}, N., et al. 2011, A\&A, 528, 16
\bibitem [Vink et al.(2001)]{2001A&A...369..574V}  {Vink}, J.~S. and {de Koter}, A. and {Lamers}, H.~J.~G.~L.~M. 2001, A\&A, 369, 574
\bibitem [Woosley et al.(1994)]{1994ApJ...429..300W} {Woosley}, S.~E., {Eastman}, R.~G., {Weaver}, T.~A., {Pinto}, P.~A. 1994, \apj, 429, 300
\bibitem [Yaron et al.(2017)]{arXiv170102596} {Yaron}, O., Perley, D. A., Gal-Yam, A., et al. 2017, arXiv:1701.02596
\bibitem [Yoon et al.(2010)]{2010ApJ...725..940Y} {Yoon}, S.-C., {Woosley}, S.~E., {Langer}, N. 2010, \apj, 725, 940
\bibitem [Yoon (2015)]{2015PASA...32...15Y} {Yoon}, S.-C. 2015, PASA, 32, 15
\bibitem [Yoon et al.(2017)]{2017arXiv170102089Y} {Yoon}, S.-C. and {Dessart}, L. and {Clocchiatti}, A. 2017, arXiv 170102089
\end{thebibliography}
\end{document}